\documentclass[twocolumn]{aastex63}
\usepackage{amssymb,amsmath}
\usepackage{graphicx}
\usepackage{xcolor,natbib}
\usepackage{multirow}
\usepackage[T1]{fontenc}
\usepackage{ae,aecompl}
\usepackage{newtxtext, newtxmath}
\usepackage{float}
\usepackage{subfigure}
\usepackage{longtable}
\usepackage{enumitem}
\usepackage{longtable,listings}
\usepackage[flushleft]{threeparttable}
\usepackage{parcolumns}
\def\oiii{[O~{\sc iii}]\ }
\def\nii{[N~{\sc ii}]\ }

\def\oii{[O~{\sc ii}]$\lambda3727$\AA\ }

\submitjournal{ApJ}

\shorttitle{Higher $n_e$ in NLRs in Type-1 AGN}
\shortauthors{ZHANG X. G.}

\begin{document}

\title{Are there higher electron densities in narrow emission line regions of Type-1 AGN than Type-2 AGN?}

\correspondingauthor{XueGuang Zhang}%
\email{xgzhang@gxu.edu.cn}
\author{XueGuang Zhang$^{*}$}
\affiliation{Guangxi Key Laboratory for Relativistic Astrophysics, School of Physical Science and Technology, 
GuangXi University, No. 100, Daxue Road, Nanning, 530004, P. R. China}

\begin{abstract}
	In the manuscript, we check properties of electron densities $n_e$ traced by flux ratio $R_{sii}$ of 
[S~{\sc ii}]$\lambda6716$\AA~ to [S~{\sc ii}]$\lambda6731$\AA~ in narrow emission line regions (NLRs) between 
Type-1 AGN and Type-2 AGN in SDSS DR12. Under the framework of Unified Model considering kpc-scale structures, 
similar $n_e$ in NLRs should be expected between Type-1 AGN and Type-2 AGN. Based on reliable measurements of 
[S~{\sc ii}] doublet with measured parameters at least five times larger than corresponding uncertainties, there 
are 6039 Type-1 AGN and 8725 Type-2 AGN (excluding the Type-2 LINERs and the composite galaxies) collected from 
SDSS DR12. Then, lower $R_{sii}$ (higher $n_e$) in NLRs can be well confirmed in Type-1 AGN than in Type-2 AGN, 
with confidence level higher than 5$\sigma$, even after considering necessary effects including effects of 
electron temperatures traced by [O~{\sc iii}]$\lambda4364,4959,5007$\AA~ on estimating $n_e$ in NLRs. Two probable 
methods are proposed to explain the higher $n_e$ in NLRs in Type-1 AGN. First, the higher $n_e$ in NLRs of Type-1 
AGN could indicate longer time durations of AGN activities in Type-1 AGN than in Type-2 AGN, if AGN activities 
triggering galactic-scale outflows leading to more electrons injecting into NLRs were accepted to explain the 
higher $n_e$ in NLRs of Type-2 AGN than HII galaxies. Second, the lower $n_e$ in NLRs of Type-2 AGN could be 
explained by stronger star-forming contributions in Type-2 AGN, considering lower $n_e$ in HII regions. The 
results provide interesting challenges to the commonly and widely accepted Unified Model of AGN.
\end{abstract}

\keywords{
galaxies:active - galaxies:nuclei - galaxies:emission lines - galaxies:Seyfert
}

\section{Introduction}

	Different observed phenomena between broad line AGN (Active Galactic Nuclei) (Type-1 AGN) and narrow line AGN 
(Type-2 AGN) can be well explained by the known Unified Model (UM) of AGN, considering expected different orientation 
angles of central accretion disk \citet{an93}, combining with different properties of central activities and inner dust 
torus etc., as well discussed in \citet{mb12, Oh15, ma16, bb18, bn19, kw21, zh22a}. More recent reviews on the UM can 
be found in \citet{nh15}. The elegant UM has been strongly supported by clearly detected polarized broad emission lines 
and/or clearly detected broad infrared emission lines in some Type-2 AGN \citep{mg90, hl97, tr03, nk04, or17, sg18, 
mb20} and strong resonance of silicate dust at 10${\rm \mu m}$ is seen in absorption towards many Type-2 AGN but in 
emission in Type-1 AGN reported in \citet{sh05}. Under the current framework of the UM, Type-1 AGN are intrinsically 
like Type-2 AGN of which central regions including central accretion power source around black hole (BH) and broad line 
regions (BLRs) are hidden by central dust torus.


	However, even after considerations of different properties of central dust torus and central activities related 
to central black hole (BH) accreting power source, there are some other challenges to the being constantly revised UM. 
\citet{fb02} have discussed the probably different evolutionary patterns in Type-1 and Type-2 AGN. \citet{hi09} have 
shown that host galaxies of Type-2 AGN have higher average star formation rates than Type-1 AGN. \citet{vk14} have 
shown different environment characteristics with different neighbours around Type-1 AGN and Type-2 AGN. More recently, 
\citet{zy19} have shown lower stellar masses of host galaxies in Type-1 AGN than Type-2 AGN, through X-ray selected AGN. 
\citet{bg20} have shown significantly different properties of UV/optical and mid-infrared colour distribution of the 
different AGN types. More recently, we \citet{zh22b} have shown statistically larger stellar velocity dispersion in 
Type-1 AGN than in Type-2 AGN. As the detailed discussions on the UM in \citet{nh15}, the UM has been successfully applied 
to explain different observed features between Type-1 and Type-2 AGN in many different ways, however, the AGN family 
with many other features considering the reported challenges to the UM are far from homogeneous.

	The UM has been well accepted that Type-1 AGN are intrinsically like Type-2 AGN, there are not only similar 
properties of central region on scale of sub-pcs including central BLRs but also similar properties of NLRs (narrow 
emission line regions) on scale of kpcs. Therefore, considering NLRs on scale of kpcs under the framework of the UM, 
there should be similar properties of electron densities in NLRs between Type-1 AGN and Type-2 AGN, which is the starting 
point of the manuscript. Moreover, not similar as properties of central power source and BLRs which can be affected by 
physical properties of dust torus on scale of pcs, there are few structures on scale of kpcs having effects on physical 
properties of NLRs on scale of kpcs. In other words, physical properties of NLRs are pure, leading to more robust final 
results without additional contaminations. Furthermore, flux ratios of [S~{\sc ii}] doublet are mainly considered in 
the manuscript, indicating that moving dust clouds and orientation effects have no effects on our final results on flux 
ratios of [S~{\sc ii}] doublet.

	Properties of electron densities $n_e$ in NLRs are mainly considered and discussed between the Type-1 AGN and 
the Type-2 AGN, which will provide further clues to support the UM or will provide further clues leading a challenge to 
the UM. Electron densities $n_e$ in emission line regions can be well and conveniently determined by narrow forbidden 
emission line ratios. In 1950s, \citet{sm54} has shown that the electron densities in planetary nebulae can be well 
estimated by relative intensities of the forbidden lines, and then followed and improved by \citet{os55, of59, os55a, 
os60, ss70, ae76}. In 1980s, \citet{cm80} have shown that forbidden [S~{\sc ii}]$\lambda6716,6731$\AA~ line ratios can 
be effectively applied to determine electron densities based on solutions of collision strengths and transition 
probabilities, and then followed by \citet{sk89}. Then, in the classic book of 'Astrophysics of Gaseous Nebulae and 
Active Galactic Nuclei' \citep{os89, of06}, there are detailed review on the theoretical method to determine electron 
densities in emission line regions by line ratios of forbidden doublets. More recently, \citet{zl13, dc14, po14, sa16, 
kn17, kg18, kn19, fm20, kn21, rr21, dv22} have shown the methods and/or corresponding discussions to determine electron 
densities in emission regions by forbidden line rations. Among line flux ratios of forbidden doublets, the ratio of 
[S~{\sc ii}]$\lambda6716,6731$\AA~ doublet is preferred in the manuscript to trace properties of $n_e$ in NLRs, because 
the collected low redshift emission line objects are from SDSS DR12 (Sloan Digital Sky Survey, Data Release 12, 
\citet{al15}), with apparent [S~{\sc ii}]$\lambda6716,6731$\AA~ doublets in their SDSS spectra.

	Based on the parameter $R_{sii}$, flux ratio of [S~{\sc ii}]$\lambda6716$\AA~ to [S~{\sc ii}]$\lambda6731$\AA, 
properties of $n_e$ in NLRs can be well conveniently checked between Type-1 AGN and Type-2 AGN collected from Sloan Digital 
Sky Survey (SDSS) data release 12 (DR12). Section 2 presents data samples of Type-1 AGN and Type-2 AGN, methods to measure 
[S~{\sc ii}] doublets. Section 3 shows main results and necessary discussions on properties of electron densities in NLRs 
of different kinds of AGN. Section 4 gives a further implication. Section 5 gives final summaries and conclusions. And in 
the manuscript, the cosmological parameters of $H_{0}~=~70{\rm km\cdot s}^{-1}{\rm Mpc}^{-1}$, $\Omega_{\Lambda}~=~0.7$ 
and $\Omega_{m}~=~0.3$ have been adopted.

\section{Data Samples}

\subsection{Parent samples of Type-1 AGN and Type-2 AGN}
	The work is based on large samples of low redshift Type-1 AGN and Type-2 AGN which have apparent 
[S~{\sc ii}]$\lambda6716,6731$\AA~ doublets. Therefore, low redshift AGN with $z~<~0.3$ in SDSS DR12 are mainly considered.

	Criterion of redshift smaller than 0.3, $z~<~0.3$, is applied to collect 12342 low redshift Type-1 AGN from SDSS 
pipeline classified QSOs \citep{rg02, ro12, pr15, lh20} in DR12, through the SDSS provided SQL (Structured Query Language) 
Search tool (\url{http://skyserver.sdss.org/dr12/en/tools/search/sql.aspx}) by the following query
\begin{lstlisting}
SELECT plate, fiberid, mjd
FROM SpecObjall
WHERE class='QSO' and z<0.30 and zwarning=0
\end{lstlisting}
where 'SpecObjall' is SDSS pipeline provided database including basic properties of emission line galaxies in SDSS DR12. 
More detailed information of the database 'SpecObjall' can be found in 
\url{http://skyserver.sdss.org/dr12/en/help/docs/tabledesc.aspx}. The collected information of plate, fiberid and mjd 
can be conveniently applied to download SDSS spectra of the 12342 Type-1 AGN.

	The same criteria $z~<~0.3$ combining with criterion 'subclass='AGN'' are applied to collect all the 16269 low 
redshift Type-2 AGN from SDSS pipeline classified main galaxies in DR12, by the following query
\begin{lstlisting}
SELECT plate, fiberid, mjd
FROM SpecObjall
WHERE class='GALAXY' and zwarning=0
   and subclass = 'AGN' and z<0.30
\end{lstlisting}
More detailed information of SDSS spectroscopic catalogs (subclass, class, etc.) can be found in 
\url{https://www.sdss.org/dr12/spectro/catalogs/}.

\subsection{Parent samples of HII galaxies}

	Besides Type-1 AGN and Type-2 AGN collected from SDSS DR12, HII galaxies are also simply discussed in the 
manuscript, which will provide clues on AGN activity contributions to properties of $n_e$ in NLRs by comparing HII 
galaxies and Type-2 AGN.

	The criterion $z~<~0.3$ combining with criterion of 'subclass='starforming'' are applied to collect all the 
245590 low redshift HII galaxies from SDSS pipeline classified main galaxies in DR12, by the following query
\begin{lstlisting}
SELECT plate, fiberid, mjd
FROM SpecObjall
WHERE class='GALAXY' and z<0.30 and zwarning=0
   and subclass = 'starforming'
\end{lstlisting}

\begin{figure*} 
\centering\includegraphics[width = 18cm,height=10cm]{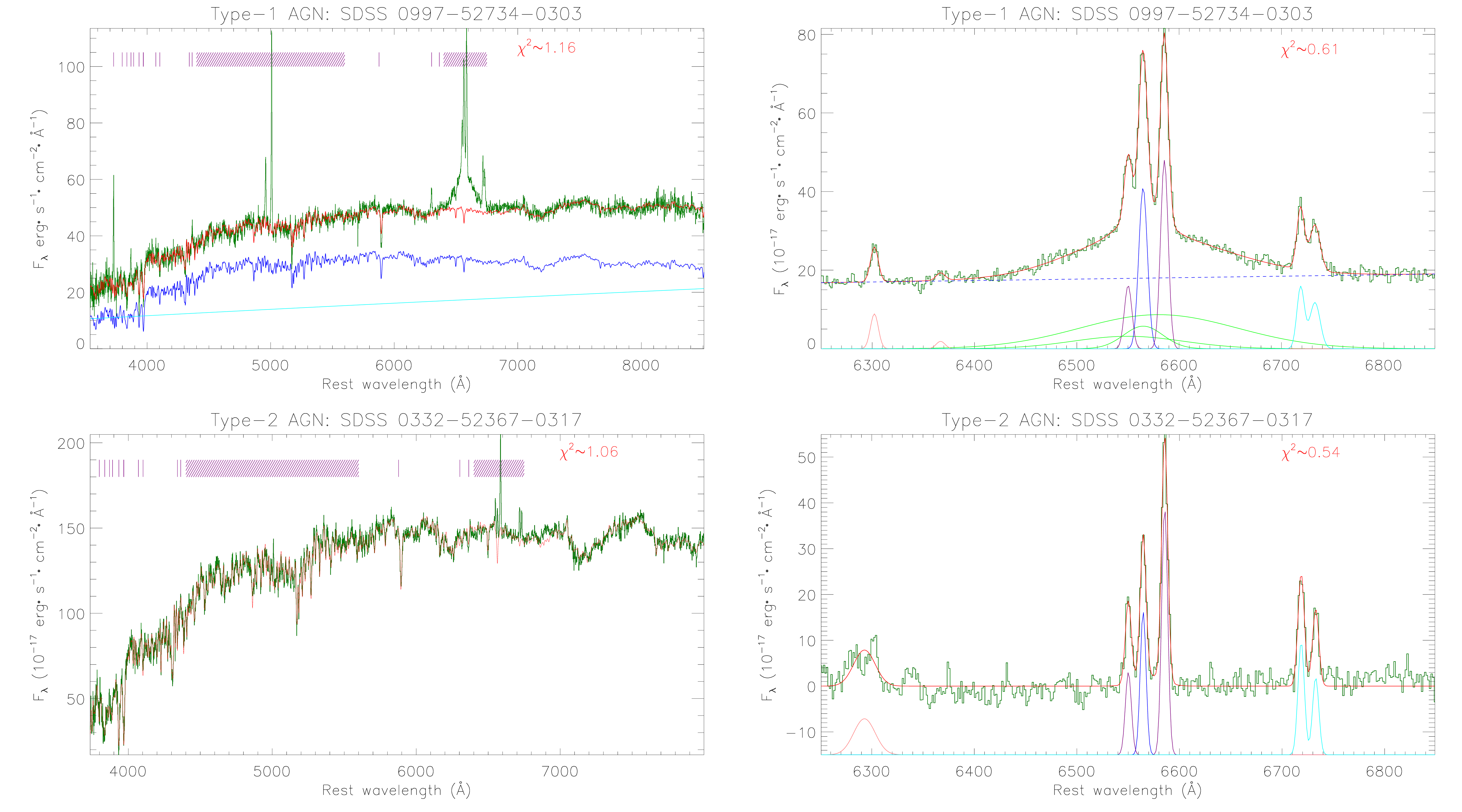}
\caption{Left panels show the SSP method determined best descriptions (solid red line) to the SDSS spectra (solid dark 
green line) of Type-1 AGN 0997-52734-0303 (PLATE-MJD-FIBERID) and Type-2 AGN 0332-52367-0317. In each left panel, from 
left to right, the vertical purple lines point out the emission lines being masked out when the SSP method is running, 
including \oii, H$\theta$, H$\eta$, [Ne~{\sc iii}]$\lambda3869$\AA, He~{\sc i}$\lambda3891$\AA, Ca~K, 
[Ne~{\sc iii}]$\lambda3968$\AA, Ca~H line, [S~{\sc ii}]$\lambda4070$\AA, H$\delta$, H$\gamma$, [O~{\sc iii}]$\lambda4364$\AA, 
He~{\sc i}$\lambda5877$\AA\ and [O~{\sc i}]$\lambda6300,~6363$\AA\ doublet, and the area filled by purple lines around 
5000\AA\ shows the region masked out including the optical Fe~{\sc ii} lines, broad and narrow H$\beta$ and \oiii doublet, 
and the area filled by purple lines around 6550\AA\ shows the region masked out including the broad and narrow H$\alpha$, 
\nii and [S~{\sc ii}] doublets. In top left panel, solid blue line shows the determined host galaxy contributions, solid 
cyan line shows the determine AGN continuum emissions. Right panels show the best descriptions (solid red line) to the 
emission lines around H$\alpha$ (solid dark green line), especially on the [S~{\sc ii}] doublet, after subtractions of 
host galaxy contributions. In each right panel, solid blue line shows the determined narrow H$\alpha$, solid purple 
lines show the determine [N~{\sc ii}] doublet, solid pink lines show the determined [O~{\sc i}] doublet, solid cyan lines 
show the determined [S~{\sc ii}] doublet. In top right panel, solid green lines show the determined broad Gaussian 
components in the broad H$\alpha$, dashed blue line shows the determined power law continuum emissions underneath the 
emission lines. In each panel, the $\chi^2$ (the summed squared residuals for the best-fitting results divided by the 
degree of freedom) is marked in red characters. 
}
\label{exams}
\end{figure*}

\begin{figure*}  
\centering\includegraphics[width = 18cm,height=6cm]{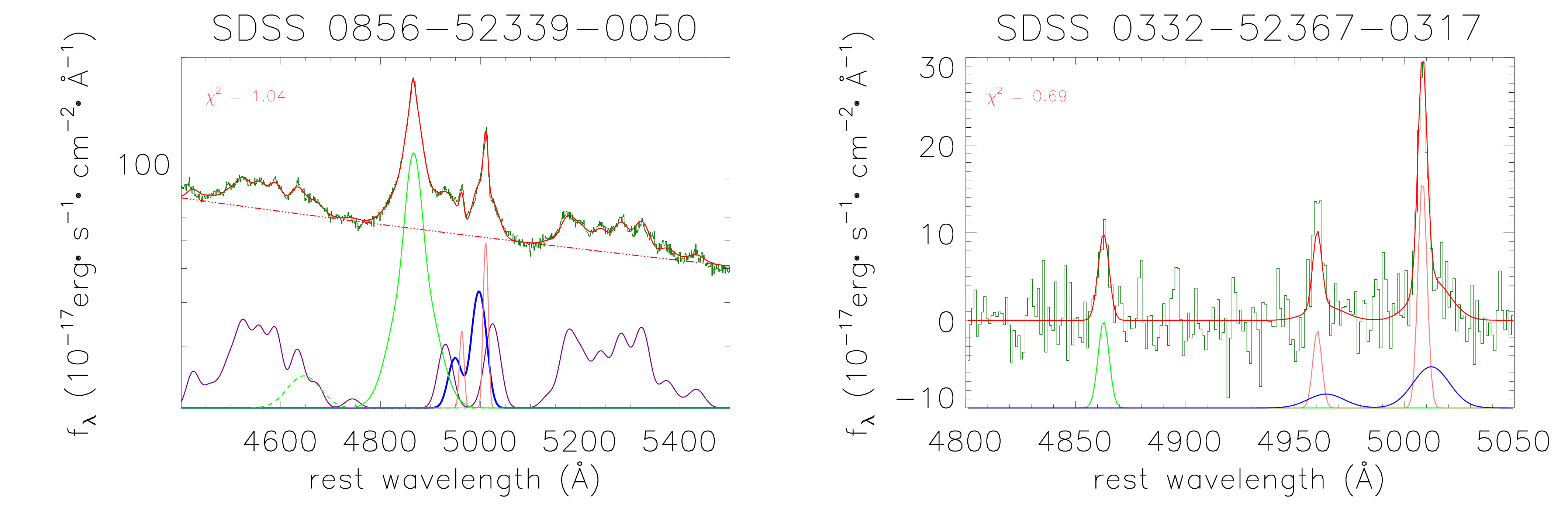}
\caption{Left panel shows the best fitting results (solid red line) to emission lines around H$\beta$ (solid dark green 
line) including apparent optical Fe~{\sc ii} emission features in the Type-1 AGN 0856-52339-0050. Double-dot-dashed red 
line shows the determined power law continuum emissions, solid green line shows the determined broad H$\beta$, solid 
purple lines show the determined optical Fe~{\sc ii} lines, dashed green line shows the determined broad He~{\sc ii} line, 
solid pink lines show the determined core \oiii components, and thick blue solid lines show the determined blue-shifted 
extended \oiii components. Right panel shows the best fitting results (solid red line) to the emission lines around 
H$\beta$ (solid dark green line) in the Type-2 AGN 0332-52367-0317, after subtractions of host galaxy contributions. 
Solid green line shows the determined narrow H$\beta$, solid lines in pink and in blue show the determined core and 
extended components of [O~{\sc iii}] doublet. And the calculated $\chi^2$ values are marked in the top-left corners 
in the panels. In order to show clearer emission features in the left panel, the Y-axis is in logarithmic coordinate.}
\label{hb}
\end{figure*}

\begin{figure*}
\centering\includegraphics[width = 18cm,height=7cm]{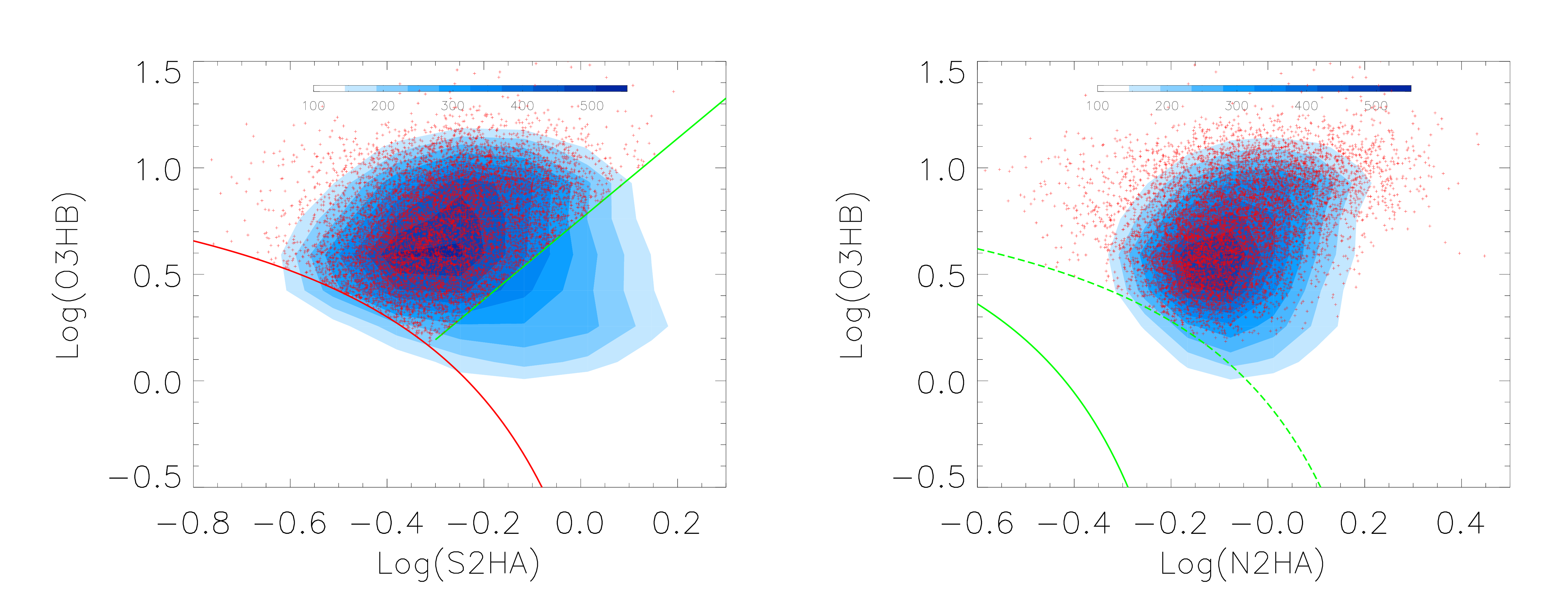}
\caption{Properties of the collected 12999 Type-2 AGN shown in contours filled by bluish colors and the 8725 Type-2 AGN 
(excluding the Type-2 LINERs and the composite galaxies) shown as red pluses in the BPT diagrams of S2HA versus O3HB 
(left panel) and of N2HA versus O3HB (right panel). In left panel, solid red line and solid green line show the dividing 
lines as discussed in \citet{kb06} between HII galaxies and AGN and between Seyfert 2 galaxies and Type-2 LINERs, leading 
Type-2 LINERs to lie into the region above the solid red line but below the solid green line. In right panel, solid and 
dashed green lines show the dividing lines between HII galaxies and composite galaxies and AGN, as discussed in \citet{ka03}.}
\label{bpt}
\end{figure*}

\begin{figure*}
\centering\includegraphics[width = 15cm,height=9cm]{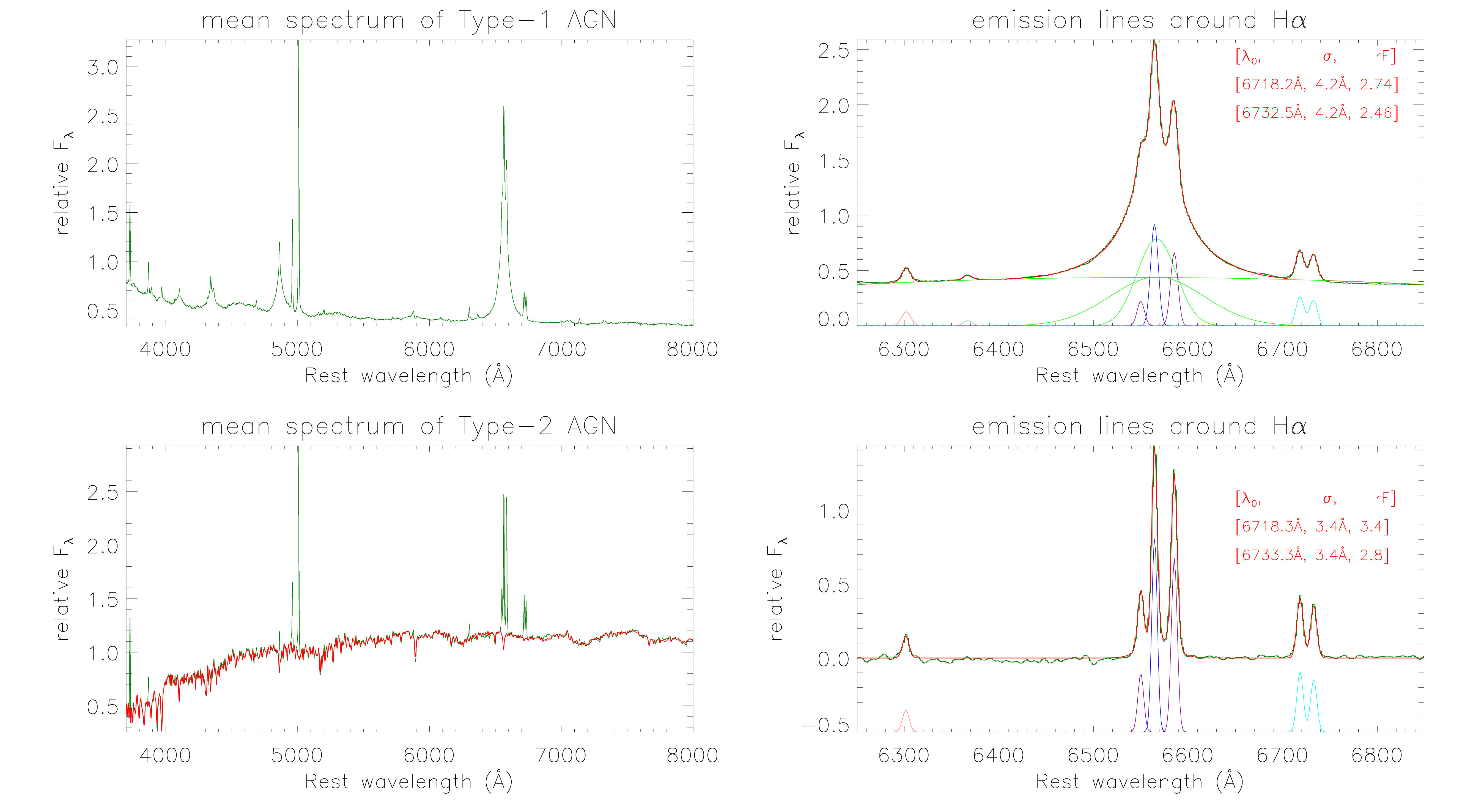}
\caption{Left panels show properties of mean spectra (in dark green) of the 1251 Type-1 AGN (top-left panel) and the 
1198 Type-2 AGN (bottom left panel) with high quality spectra in the main samples. In bottom left panel, solid red line 
shows the SSP method determined host galaxy contributions. Right panels show the best fitting results to the emission 
lines around H$\alpha$ in the mean spectrum of Type-1 AGN (top-right panel) and of Type-2 AGN after subtractions of 
the host galaxy contributions (bottom-right panel). In right panels, the symbols and line styles are the same as those 
in right panels of Fig.~\ref{exams}. In each right panel, top right corner lists the measured line parameters [central 
wavelength $\lambda_0$, second moment $\sigma$, relative flux $rF$] of the [S~{\sc ii}] doublet.}
\label{msp}
\end{figure*}

\subsection{Method to measure the line parameters of [S~{\sc ii}] doublet}


	In order to well measure line intensities of [S~{\sc ii}]$\lambda6716,6731$\AA, host galaxy contributions 
included in SDSS spectra should be firstly subtracted.

	The common SSP method (Simple Stellar Population) \citep{bc03, ka03, cm05, cm17} has been applied to determine 
the host galaxy contributions, similar as what we have done in \citet{zh14, zh16, ra17, zh19, zh21m, zh21a, zh21b, 
zh22b, zh23}. We have exploited the 39 simple stellar population templates in \citet{bc03}, which can be used to well 
describe the characteristics of almost all the SDSS galaxies as discussed in \citet{bc03}. Meanwhile, an additional 
power law component is applied to describe intrinsic AGN continuum emissions, especially in Type-1 AGN. Meanwhile, 
when the SSP method is running, narrow emission lines in spectrum are masked out, by full width at zero intensity 
about 450${\rm km~s^{-1}}$. And the wavelength ranges from 4450 to 5600\AA~ and from 6250 to 6750\AA~ are also masked 
out for the broad H$\beta$ and the broad H$\alpha$ and optical Fe~{\sc ii} emission lines. Then, through the 
Levenberg-Marquardt least-squares minimization technique (the known MPFIT package), the best descriptions can be well 
determined to the SDSS spectra with emission lines being masked out. Moreover, when the SSP method is running, only 
one restriction is accepted that the strengthened factor of each stellar population template is not smaller than zero. 
Left panels of Fig.~\ref{exams} shows two examples on the SSP method determined host galaxy contributions in one 
Type-1 AGN and one Type-2 AGN.

	After subtractions of the host galaxy contributions (if there are), emission lines around H$\alpha$, within 
rest wavelength from 6250 to 6850\AA, can be well described by multiple Gaussian functions. Simple descriptions on the 
measurements of emission lines are as follows, similar as what we have recently done in \citet{zh21a, zh21b, zh21c}. 
Three broad Gaussian functions plus one narrow Gaussian function are applied to describe the broad and narrow H$\alpha$, 
six narrow Gaussian components are applied to describe the [O~{\sc i}], [N~{\sc ii}] and [S~{\sc ii}] doublets, a 
power law component is applied to describe the continuum emissions underneath the broad H$\alpha$. Then, through 
the Levenberg-Marquardt least-squares minimization technique, emission lines can be well described by multiple 
Gaussian functions, and uncertainties (formal $1\sigma$ errors) of the model parameters can be determined by the 
covariance matrix. When the model functions above are applied, the following restrictions are accepted. First, the 
components of each forbidden narrow emission line doublet ([S~{\sc ii}], [N~{\sc ii}], [O~{\sc i}]) have the same 
redshift and the same line width in velocity space. Second, each emission component has intensity not smaller than 
zero. Third, each narrow Gaussian component has line width (second moment) smaller than $500{\rm km/s}$\footnote{The 
maximum full width at half maximum of narrow H$\alpha$ of SDSS quasars in \citet{sh11} is around 1200${\rm km/s}$ 
(second moment about $509{\rm km/s}$). Therefore, $500{\rm km/s}$ is accepted as the upper limit of second moment of 
narrow emission lines.}. Fourth, each broad Gaussian component in broad H$\alpha$ has line width larger than the 
line width of narrow H$\alpha$. Fifth, the flux ratio of [N~{\sc ii}] doublet is fixed to the theoretical value 3. 
And when the fitting procedure is running, the starting values of the parameters are as follows. For each narrow 
emission line, theoretical central wavelength, 2\AA~ and 0 are accepted the starting values of central wavelength, 
second moment and line intensity. For the three broad Gaussian components in broad H$\alpha$, the starting values 
of [central wavelength, second moment, intensity] are [6540, 20, 0], [6564, 25, 0] and [6580, 20, 0], respectively. 
Right panels of Fig.~\ref{exams} shows two examples on the best descriptions to the emission lines around H$\alpha$, 
after subtractions of host galaxy contributions. 

	Moreover, because line intensities of [O~{\sc iii}]$\lambda4959, 5007$\AA~ and narrow H$\beta$ will be 
discussed in the following section, simple descriptions are as follows on the fitting procedure applied to describe 
the emission lines around H$\beta$ within rest wavelength range from 4400 to 5600\AA~ after subtractions of the host 
galaxy contributions. Similar as what we have recently done in \citet{zh21a, zh21b, zh21c}, three broad Gaussian 
functions plus one narrow Gaussian function are applied to describe the broad and narrow H$\beta$, two narrow and 
two broad Gaussian components are applied to describe the core and extended components of 
[O~{\sc iii}]$\lambda4959,5007$\AA~ doublet \citep{sh11, gh05}, one Gaussian component is applied to describe the 
He~{\sc ii} line, broadened and scaled Fe~{\sc ii} templates discussed in \citet{kp10} are applied to describe 
optical Fe~{\sc ii} lines, and a power law component is applied to describe the continuum emissions underneath the 
broad H$\beta$. The following restrictions are accepted to the model parameters, as the restrictions to the model 
parameters to describe the emission lines around H$\alpha$. First, the core (extended) components of [O~{\sc iii}] 
doublet have the same redshift, the same line width and the flux ratio fixed to the theoretical value 3. Second, 
each emission component has intensity not smaller than zero. Third, the core components of [O~{\sc iii}] doublet 
and the narrow H$\beta$ have line widths (second moment) smaller than $500{\rm km/s}$. Fourth, each broad Gaussian 
component in broad H$\beta$ has line width larger than the line width of narrow H$\beta$. Fifth, the extended 
components of [O~{\sc iii}] doublet have line widths larger than the line widths of the core components. And when 
the fitting procedure is running, the starting values of the parameters are as follows. For each narrow emission lines, 
theoretical central wavelength, 2\AA~ and 0 are accepted the starting values of central wavelength, second moment and 
line intensity. For the three broad Gaussian components in broad H$\beta$, the starting values of [central wavelength, 
second moment, intensity] are [4840, 20, 0], [4861, 25, 0] and [4880, 20, 0], respectively. Fig.~\ref{hb} shows two 
examples on the best-fitting results to the emission lines around H$\beta$ in one Type-1 AGN and one Type-2 AGN, 
through the Levenberg-Marquardt least-squares minimization technique.

\begin{figure}  
\centering\includegraphics[width = 8cm,height=10cm]{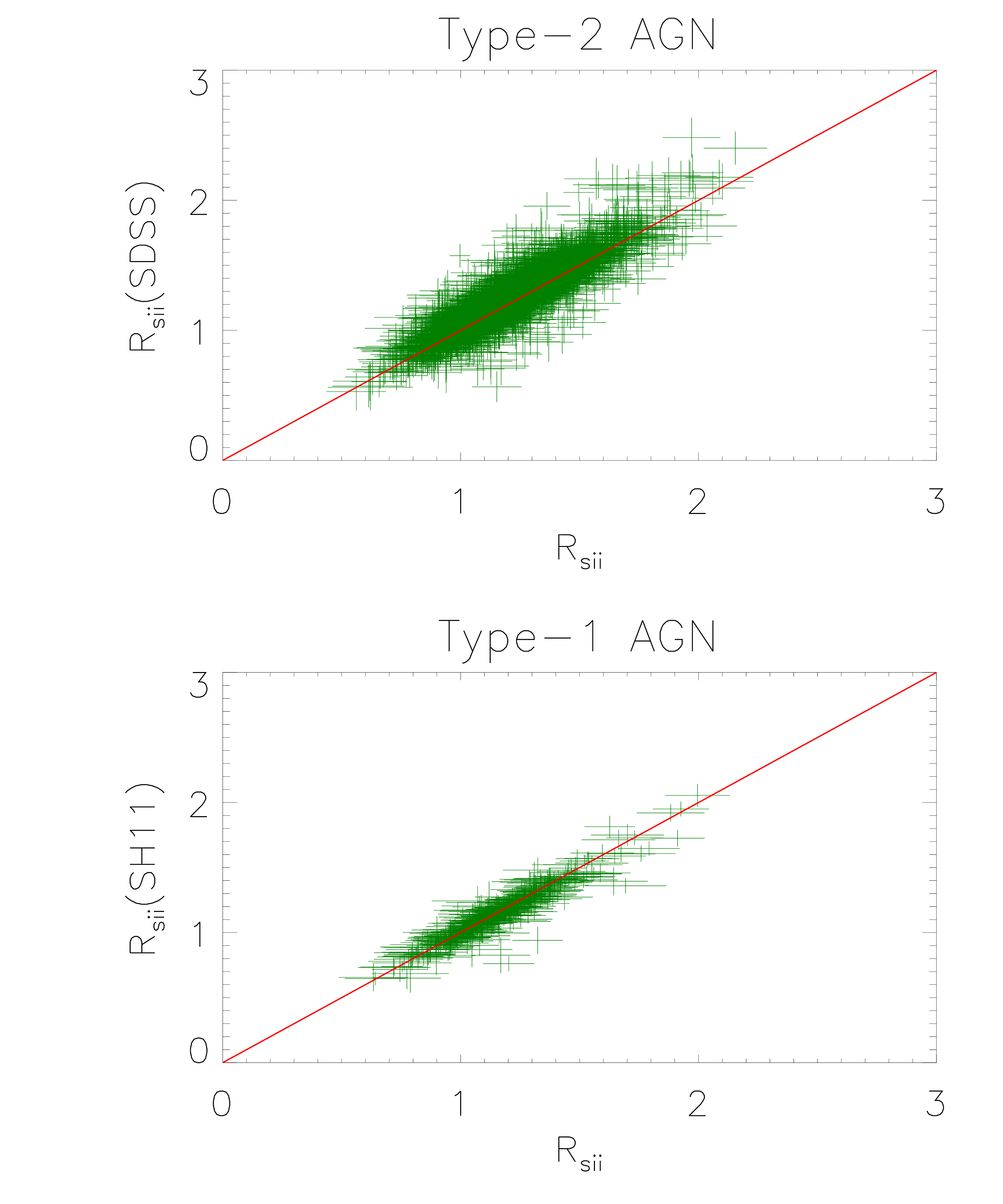}
\caption{On the correlations between measured $R_{sii}$ in the manuscript and $R_{sii}(SDSS)$ determined from the 
SDSS pipeline determined line parameters of the Type-2 AGN (top panel), and between the measured $R_{sii}$ in the 
manuscript and $R_{sii}(SH11)$ determined from the reported line parameters of the Type-1 AGN in \citet{sh11} 
(bottom panel). In each panel, solid red line shows $X~=~Y$.
}
\label{comp}
\end{figure}

\begin{figure} 
\centering\includegraphics[width = 8cm,height=5.5cm]{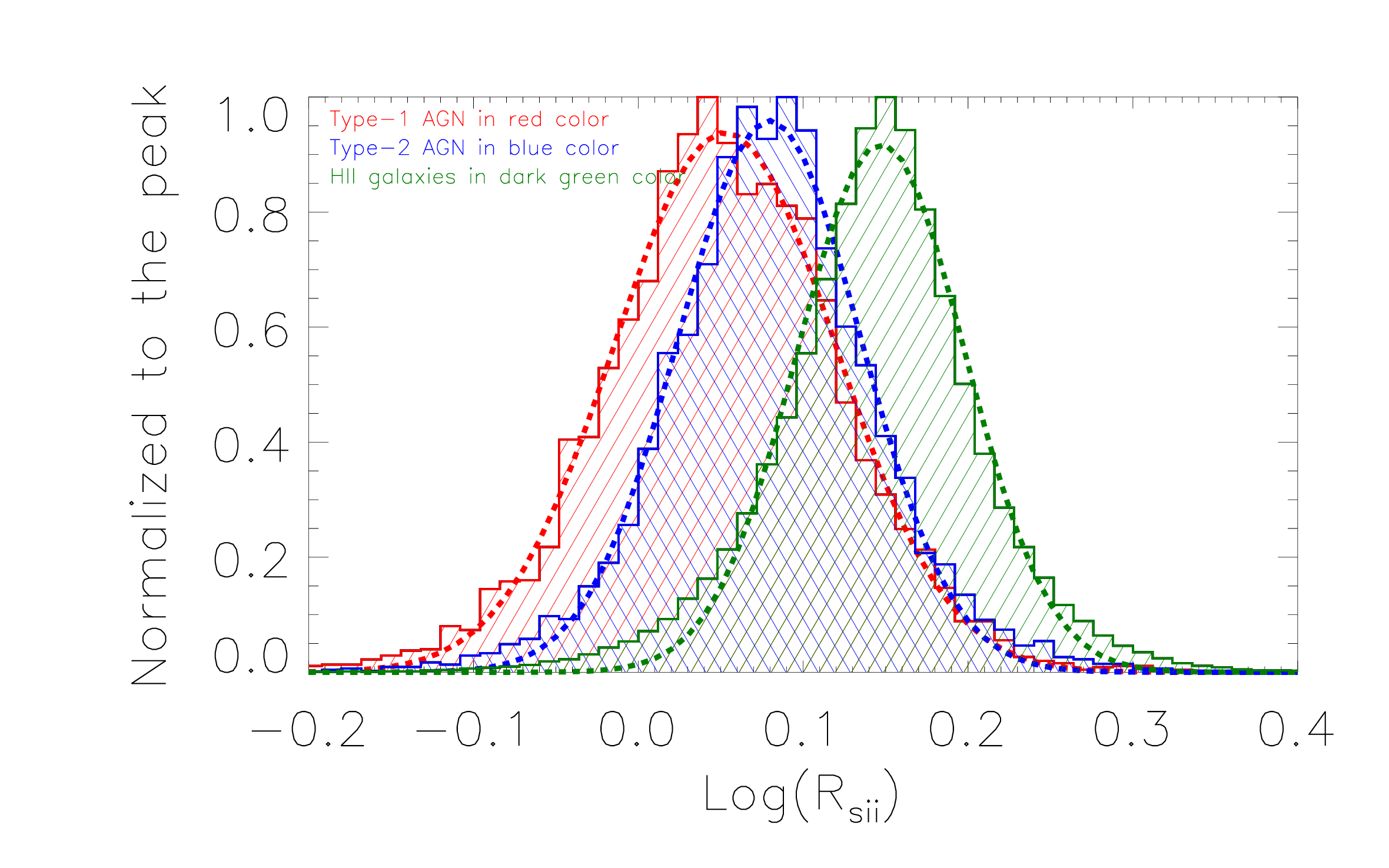}
\caption{Distributions of $\log(R_{sii})$ of the 6039 Type-1 AGN (histogram filled by red lines), the 8725 Type-2 AGN 
(histogram filled by blue lines), and the 199700 HII galaxies (histogram filled by dark green lines) in the final 
main samples, respectively. Thick dashed lines in red, in blue and in dark represent the corresponding best Gaussian 
profiles for the $\log(R_{sii})$ distributions of the Type-1 AGN, the Type-2 AGN and the HII galaxies, respectively.
}
\label{dis}
\end{figure}

\begin{figure*} 
\centering\includegraphics[width = 18cm,height=9cm]{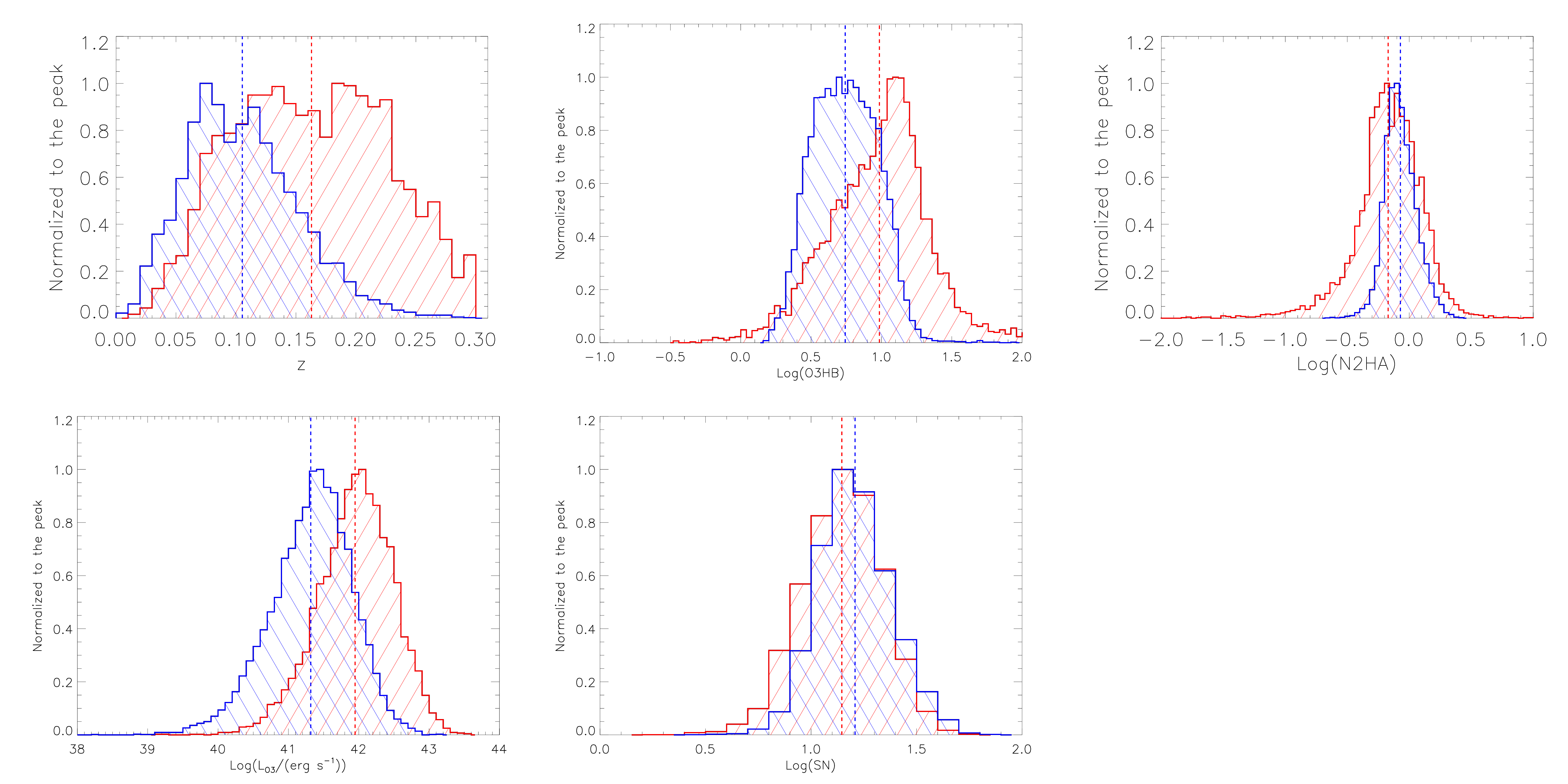}
\caption{Distributions of redshift, O3HB, N2HA, $L_{O3}$ and SN of the Type-1 AGN (histogram filled with red lines) 
and Type-2 AGN (histogram filled with blue lines) in the main samples. In each panel, vertical dashed line in red 
and in blue mark position of mean value of each distribution of Type-1 AGN and Type-2 AGN, respectively.
}
\label{d1p}
\end{figure*}

\begin{figure*} 
\centering\includegraphics[width = 18cm,height=9cm]{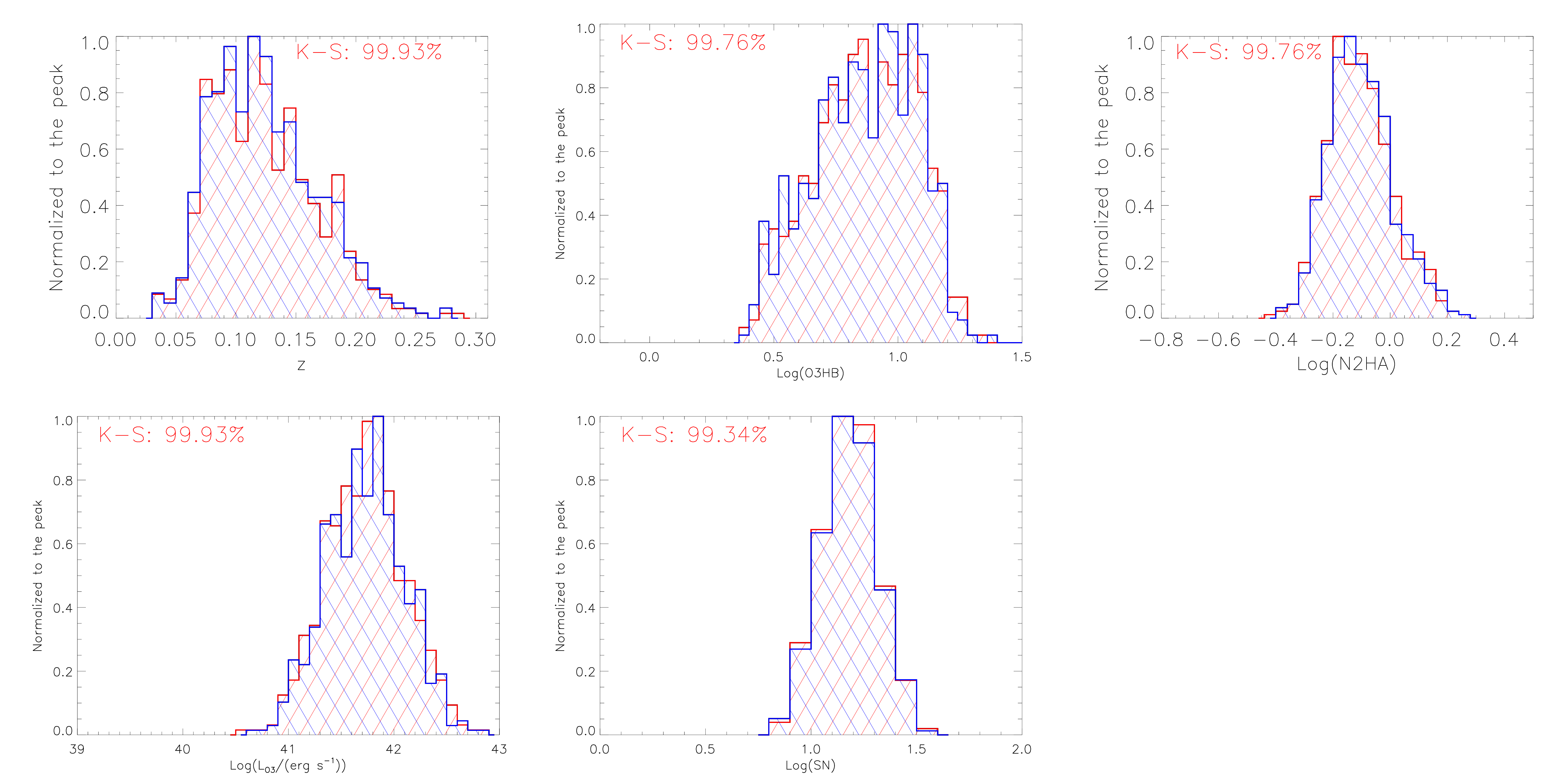}
\caption{Distributions of redshift, O3HB, N2HA, $L_{O3}$ and SN of the 548 Type-1 AGN (histogram filled with red lines) 
and the 548 Type-2 AGN (histogram filled with blue lines) in the subsamples. In each panel, the Kolmogorov-Smirnov 
statistic technique provided significance level is marked in red characters.
}
\label{d2p}
\end{figure*}

\subsection{Final main data samples}
 
	Finally, starting with 12342 Type-1 AGN in the parent sample collected from SDSS pipeline classified quasars, 
and with 16269 Type-2 AGN in the parent sample collected from SDSS pipeline classified main galaxies, and 245590 HII 
galaxies in the parent sample collected from SDSS pipeline classified main galaxies, applying the following criteria,
\begin{itemize}
\item The measured line width and line flux of [S~{\sc ii}] doublet described by Gaussian functions are at least 5 
	times larger than their corresponding uncertainties, indicating reliable [S~{\sc ii}] doublet.
\item For the Type-1 AGN, not only there are reliable [S~{\sc ii}] doublet, but also there are reliable broad H$\alpha$ 
	emission lines with at least one broad Gaussian component with the measured line flux and line width at least 
	5 times larger than the corresponding uncertainties and second moment larger than 600${\rm km\cdot~s^{-1}}$.
\item For the Type-2 AGN, not only there are reliable [S~{\sc ii}] doublet, but also there are no broad H$\alpha$ 
	emission lines with the determined three broad Gaussian components for broad H$\alpha$ with the measured line 
	fluxes and line widths 2 times smaller than the corresponding uncertainties.
\item For the HII galaxies, not only there are reliable [S~{\sc ii}] doublet, but also there are no broad H$\alpha$ 
	emission lines with the determined three broad Gaussian components for broad H$\alpha$ with the measured line 
	fluxes and line widths 2 times smaller than the corresponding uncertainties.
\end{itemize}
leads main samples including 6039 Type-1 AGN with both apparent [S~{\sc ii}] doublets and apparent broad H$\alpha$ 
emission lines, and 12999 Type-2 AGN with apparent [S~{\sc ii}] doublets but no broad H$\alpha$ emission lines, and 
199700 HII galaxies with apparent [S~{\sc ii}] doublets but no broad H$\alpha$ emission lines. Here, the word "reliable" 
means the Gaussian function described emission component has its measured line parameters (central wavelength, 
second moment and line intensity) at least 5 times larger than the corresponding uncertainties.

	Furthermore, as described in subsection 2.1, both Seyfert 2 galaxies and Type-2 LINERs (Low Ionization Nuclear 
Emission Line Regions) (LINERs without apparent broad emission lines) are collected into the main sample of Type-2 AGN. 
However, not similar as Seyfert 2 galaxies totally powered by central BH accreting process,there are different mechanisms 
applied to Type-2 LINERs, such as shock heating \citep{ht80, ds96}, photoionization by young stars \citep{tm85, ft92}, 
photoionization by post-asymptotic giant branch (post-AGB) stars \citep{eh10, csm11}, etc. More recent review on LINERs 
can be found in \citet{mm17} which have shown that 60\% to 90\% of LINERs could be well considered as genuine AGN. 
Considering the controversial conclusion on physical nature of Type-2 LINERs (at least part of Type-2 AGN without AGN 
nature), Type-2 LINERs are not considered in the manuscript, in order to ignore effects of different physical natures 
of part of Type-2 LINERs on our final results. Not similar as Type-2 LINERs, Type-1 LINERs (LINERs with apparent broad 
emission lines) included in the parent sample of Type-1 AGN are well considered as AGN, due to their broad emission lines.

        Based on the dividing lines between Seyfert 2 galaxies and Type-2 LINERs in the BPT diagram of O3HB (flux ratio 
of [O~{\sc iii}]$\lambda5007$\AA~ to narrow H$\beta$) versus S2HA (flux ratio of total [S~{\sc ii}]$\lambda6716,6731$\AA~ 
to narrow H$\alpha$) as shown in \citet{kb06}
\begin{equation}
\begin{split}
\log(O3HB)~&>~\frac{0.72}{\log(S2HA)-0.32}+1.30\\ 
\log(O3HB)~&>~1.89\log(S2HA) +0.76
\end{split}
\end{equation}
there are 8793 Type-2 AGN, excluding the Type-2 LINERs and excluding the classified HII galaxies in the BPT diagrams 
of O3HB versus S2HA. Meanwhile, based on the dividing line between AGN and composite galaxies as discussed in \citet{ka03} 
in the BPT diagram of O3HB versus N2HA (flux ratio of [N~{\sc ii}]$\lambda6583$\AA~ to narrow H$\alpha$)
\begin{equation}
\log(O3HB)~>~\frac{0.61}{\log(N2HA)-0.47}+1.19
\end{equation}
among the 8793 Type-2 AGN, there are 68 classified composite galaxies excluded from the collected Type-2 AGN, in order 
to ignore probable strong effects of starforming. Therefore, there are 8725 (8793-68) Type-2 AGN included in the final 
main sample of Type-2 AGN. Fig.~\ref{bpt} shows properties of the collected Type-2 AGN in the BPT diagrams of S2HA 
versus O3HB (left panel) and of N2HA versus O3HB (right panel). The results in left panel of Fig.~\ref{bpt} show clear 
classifications of Type-2 LINERs. And the results in right panel of Fig.~\ref{bpt} show clear evidence to support that 
the collected Type-2 AGN, neither including Type-2 LINERs nor including composite galaxies, are reliable AGN with 
central AGN activities.

\subsection{Spectroscopic properties of mean spectra of Type-1 AGN and Type-2 AGN}

	In the subsection, mean spectra are discussed in Type-1 AGN and Type-2 AGN, in order not only to provide further 
evidence to support that the emission line fitting procedure is appropriate and but also to provide further clues to 
answer the question whether asymmetric line profiles should be considered in [S~{\sc ii}] doublet.

	The commonly accepted PCA (Principal Component Analysis) technique is applied to create mean spectra of Type-1 
AGN and Type-2 AGN. PCA technique uses an orthogonal transformation to convert a set of observations of possibly correlated 
variables into a set of values of uncorrelated variables called principal components. Commonly, mean subtraction (or mean 
centering) is necessary for performing PCA to ensure that the first principal component describes the direction of maximum 
variance. However, if mean subtraction is not performed, the PCA technique determined first eigencomponent represents the 
mean spectrum of input set of spectra. Here, we apply the convenient and public IDL PCA program pca\_solve.pro written 
by D. Schlegel, which is included in SDSS software package of IDLSPEC2D (\url{http://spectro.princeton.edu/}).

	Here, in order to check probable asymmetric profiles of [S~{\sc ii}] doublet, 1251 Type-1 AGN with spectral 
signal-to-noise larger than 20 and 1198 Type-2 AGN with spectral signal-to-noise larger than 25 are mainly considered. 
PCA technique determined mean spectra are shown in left panels of Fig.~\ref{msp}. The same SSP method is applied to 
determine host galaxy contributions in the mean spectrum of Type-2 AGN. Then, the same emission line fitting procedure 
discussed in subsection 2.2 is applied to measure the emission lines around H$\alpha$ in the mean spectra of Type-1 
AGN and of Type-2 AGN after subtractions of host galaxy contributions. The best fitting results are shown in right 
panels of Fig.~\ref{msp} to the emission lines around H$\alpha$, with the determined line parameters of [S~{\sc ii}] 
doublet marked in top right corner in each right panel.

	It is clear that the two Gaussian components can be well applied to describe the [S~{\sc ii}] doublet in the 
mean spectra of high quality Type-1 AGN and high quality Type-2 AGN, indicating there are few contributions of 
asymmetric kinematic components in [S~{\sc ii}] doublets. Therefore, the results in Fig.~\ref{msp} not only can be 
applied to support that the emission line fitting procedure can be well accepted, but also can be applied to support 
that there are few effects of asymmetric kinematic components in [S~{\sc ii}] doublets on our final results.

\begin{figure*}
\centering\includegraphics[width = 18cm,height=15cm]{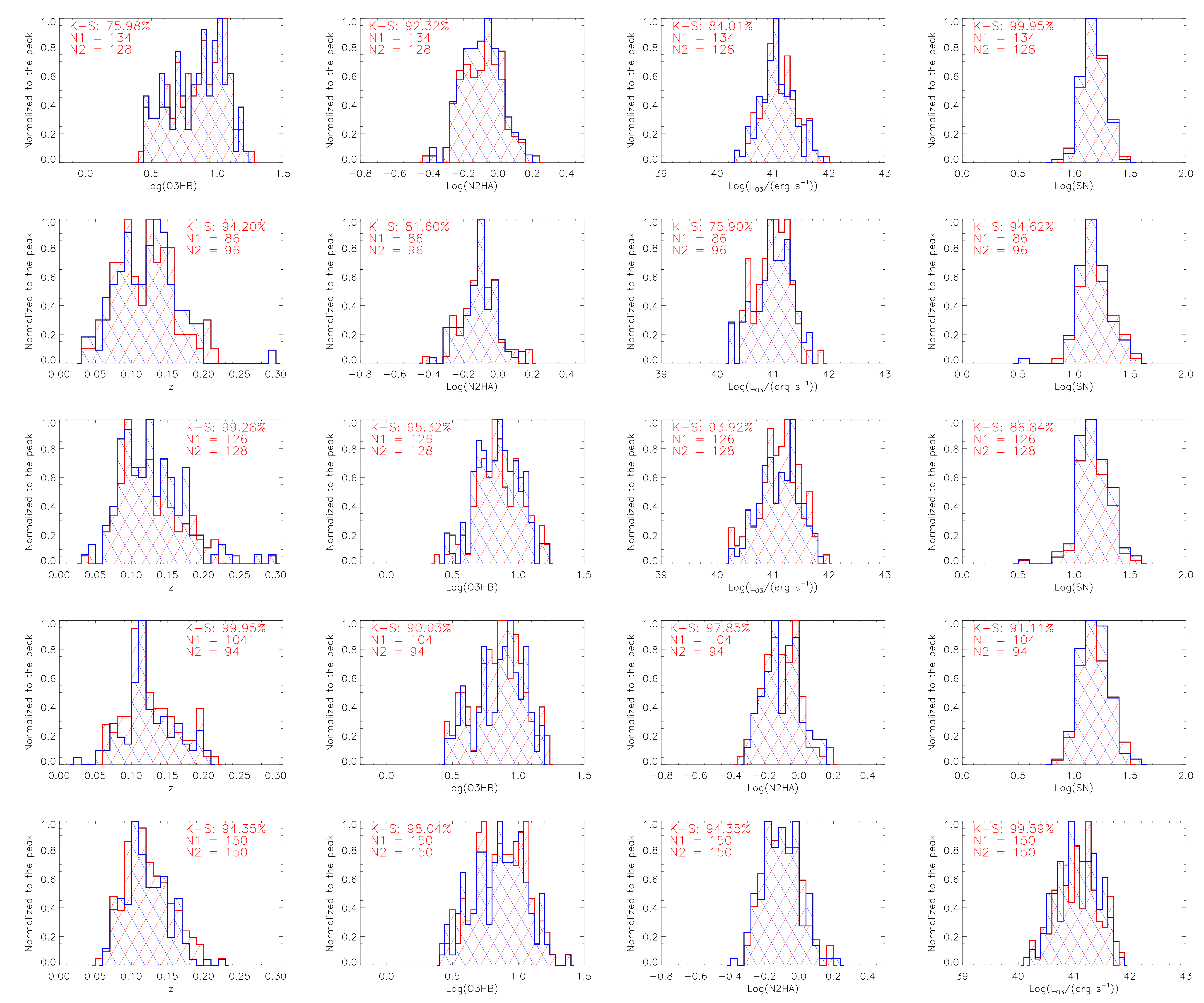}
\caption{Within a narrow range of one parameter, distributions of the other four parameters of the collected Type-1 
AGN and Type-2 AGN from the subsamples. Symbols and line styles have the same meanings as those in Fig.~\ref{d2p}. 
And the numbers N1 and N2 of the collected Type-1 AGN and Type-2 AGN are marked in red characters in top region in 
each panel. From top to bottom, the Type-1 and Type-2 AGN are collected from the subsamples, through the criteria that 
$|z-\overline{z}|~<\sim0.0134$, $|\log(O3HB)-\overline{\log(O3HB)}|~<~\sim0.051$, 
$|\log(N2HA)-\overline{\log(N2HA)}|~<~\sim0.036$, $|\log(L_{O3})-\overline{\log(L_{O3})}|~<~0.092$, 
$|\log(SN)-\overline{\log(SN)}|~<~0.05$, where $\overline{p}$ as mean value of parameter $p$. 
}
\label{d3s}
\end{figure*}

\section{Main Results and Discussions}

\subsection{To confirm the reliability of the measured line parameters of [S~{\sc ii}] doublet}

	Comparing with line parameters from different methods/techniques can provide further and necessary information 
to confirm the reliability of the measured line parameters. For the [S~{\sc ii}] doublet of which features with few 
effects from host galaxy absorption features in Type-2 AGN, it is necessary and interesting to confirm the reliability 
of our measured parameters of [S~{\sc ii}] doublet, through comparing our measured values and the values calculated 
from SDSS pipeline provided parameters for the Type-2 AGN. Due to apparent effects of broad H$\alpha$ on measured line 
parameters of [S~{\sc ii}] doublet in Type-1 AGN, the parameters reported in \citet{sh11} rather than the parameters 
reported by the SDSS pipeline are considered for the Type-1 AGN. 

	In the subsection, properties of line flux ratio $R_{sii}$ of [S~{\sc ii}]$\lambda6716$\AA~ to 
[S~{\sc ii}]$\lambda6731$\AA~ are well discussed, based on the SDSS pipeline produced line parameters of the Type-2 AGN, 
and based on the line parameters reported in \citet{sh11} of the Type-1 AGN in SDSS DR7 (Data Release 7).

	For the 8725 Type-2 AGN in SDSS DR12 (excluding the Type-2 LINERs and composite galaxies), the SDSS pipeline 
measured line parameters of [S~{\sc ii}] doublets are stored in the database of 'galSpecLine'\footnote{Detailed information 
of 'galSpecLine' can be found in \url{http://skyserver.sdss.org/dr12/en/help/docs/tabledesc.aspx}}. Top panel of 
Fig.~\ref{comp} shows the correlation between the measured $R_{sii}$ in the manuscript and $R_{sii}(SDSS)$ determined 
from the SDSS reported line parameters. There is a strong positive linear correlation with Spearman Rank correlation 
coefficient about 0.935 with $P_{null}~<~10^{-20}$. The linear correlation can be described by 
\begin{equation}
\begin{split}
\log(R_{sii}(SDSS))~=~&(-0.053\pm0.008)\\
	&~+~(1.059\pm0.007)\log(R_{sii})
\end{split}
\end{equation}
after considering the uncertainties in both coordinates, through FITEXY 
code\footnote{\url{https://idlastro.gsfc.nasa.gov/ftp/pro/math/fitexy.pro}}. And the mean ratio of $R_{sii}$ to 
$R_{sii}(SDSS)$ is about 0.981$\pm$0.001. The uncertainty 0.001 is determined by the commonly applied bootstrap method 
within 1000 loops. For each loop, more than half of the data points in the sample of ratio of $R_{sii}$ to $R_{sii}(SDSS)$ 
are randomly collected to create a new sample, leading to a re-measured mean value. After 1000 loops, based on distribution 
of the 1000 re-measured mean values, half width at half maximum is accepted as the uncertainty of the mean value. Therefore, 
the results in top panel of Fig.~\ref{comp} can provide strong evidence to confirm the reliability of the measured line 
parameters of [S~{\sc ii}] doublets of the Type-2 AGN in the main sample.

	Meanwhile, bottom panel of Fig.\ref{comp} shows the measured $R_{sii}$ in the manuscript and $R_{sii}(SH11)$ 
determined from the reported line parameters in \citet{sh11} for the Type-1 AGN in SDSS DR7. There is a strong positive 
linear correlation with Spearman Rank correlation coefficient about 0.96 with $P_{null}~<~10^{-20}$. After considering 
the uncertainties in both coordinates, through the FITEXY code, the linear correlation can be described by
\begin{equation}
\begin{split}
\log(R_{sii}(SH11))~=~&(-0.022\pm0.029)\\
	&~+~(1.012\pm0.025)\log(R_{sii})
\end{split}
\end{equation}
And the mean ratio of $R_{sii}$ to $R_{sii}(SH11)$ is about 1.007$\pm$0.006, with uncertainty determined by the bootstrap 
method within 1000 loops applied to data sample of the ratio of $R_{sii}$ to $R_{sii}(SH11)$. Therefore, the results in 
bottom panel of Fig.~\ref{comp} can provide strong evidence to confirm the reliability the measured line parameters of 
[S~{\sc ii}] doublets of the Type-1 AGN in the main sample.

\subsection{Direct comparisons of $R_{sii}$}

	Based on the reliable measurements of [S~{\sc ii}] doublets, distributions of $R_{sii}$ (flux ratio of 
[S~{\sc ii}]$\lambda6716$\AA~ to [S~{\sc ii}]$\lambda6731$\AA) are shown in Fig.~\ref{dis} of the Type-1 AGN, the 
Type-2 AGN and the HII galaxies in the main samples. Mean values of $\log(R_{sii})$ are about 0.052$\pm$0.005, 
0.081$\pm$0.003 and 0.147$\pm$0.001 of the Type-1 AGN, the Type-2 AGN (excluding the Type-2 LINERs and composite 
galaxies) and the HII galaxies, respectively. Uncertainty of each mean value is determined by the bootstrap method 
with 1000 loops applied. Meanwhile, based on the measured [S~{\sc ii}] doublets in the mean spectra of high quality 
Type-1 AGN and high quality Type-2 AGN in Fig.~\ref{msp}, $\log(R_{sii})$ are about 0.047 and 0.084\ in high quality 
Type-1 AGN and in high quality Type-2 AGN, respectively, which are a bit different from the mean values in the AGN 
in the main samples, indicating a few effects of spectral signal-to-noise (SN) on our final results, besides to show 
the different $\log(R_{sii})$ between high quality Type-2 AGN and high quality Type-1 AGN.

	Based on the theoretical dependence of $n_e$ on $R_{sii}$ more recently discussed in \citet{sa16, kn19},
\begin{equation}
\frac{n_e}{\rm cm^{3}}~=~\frac{627.1~\times~R_{sii}~-~909.17}{0.4315~-~R_{sii}}
\end{equation}
the mean electron densities $n_e$ can be roughly estimated as 291$\pm$18, 198$\pm$4 and 30$\pm$3\ in units of ${\rm cm^{-3}}$ 
of the 6039 Type-1 AGN, the 8725 Type-2 AGN and the HII galaxies, respectively, with uncertainties determined by accepted 
corresponding uncertainties of $R_{sii}$. Here, as discussed results in \citet{sa16, kn19}, effects of electron 
temperature on $n_e$ can lead to about 15\% uncertainties of $n_e$, which cannot be applied to explain the apparent difference 
in $n_e$ in the different kinds of emission line objects. And detailed discussions on effects of electron temperatures on 
estimating electron densities in NLRs can be found in the following subsection 3.6.

	Moreover, as discussed in \citet{kn17}, $R_{sii}$ should be effectively limited to the range from 0.4 to 1.5, 
when $R_{sii}$ is applied to calculate electron density $n_e$. Then, with $R_{sii}$ larger than 0.4 and smaller than 1.5, 
mean values of $\log(R_{sii})$ are 0.044$\pm$0.003, 0.071$\pm$0.003 and 0.116$\pm$0.001, and the corresponding mean $n_e$ 
in unit of ${\rm cm^{-3}}$ can be estimated as 319$\pm$11, 229$\pm$4 and 102$\pm$3 of the 5467 Type-1 AGN, the 8389 Type-2 
AGN and the 144210 HII galaxies among the objects in the main samples, respectively. The results can also roughly lead to 
apparently lower $\log(R_{sii})$ (higher $n_e$) in NLRs in Type-1 AGN, before considering necessary effects 
on the $R_{sii}$ comparisons between the Type-1 AGN and the Type-2 AGN.

	Considering the effective range of $R_{sii}$ to estimate $n_e$ in NLRs, the following discussed main samples of 
AGN include the 5467 Type-1 AGN with $0.4~<~R_{sii}~<~1.5$ and the 8389 Type-2 AGN with $0.4~<~R_{sii}~<~1.5$.

\subsection{Effects of different distributions of redshift, O3HB, N2HA or [O~{\sc iii}] line luminosity?}

	In order to well explain the determined apparently higher $n_e$ (only related to lower $R_{sii}$) in 
NLRs in Type-1 AGN than in Type-2 AGN which are against the expected results by the Unified model of AGN, different effects 
are considered as follows, especially based on the different distributions of redshift, O3HB, N2HA and [O~{\sc iii}] line 
luminosity $L_{O3}$ between the 5467 Type-1 AGN and the 8389 Type-2 AGN in the main samples with $0.4~<~R_{sii}~<~1.5$. 
Distributions of the parameters of redshift, O3HB, N2HA, $L_{O3}$ and SN are shown in Fig.~\ref{d1p}. Here, redshift can 
be considered as evolutionary histories of AGN. And O3HB and N2HA can be well applied in BPT diagram \citep{bpt, kb01, ka03, 
kb06, kb19, zh20} to identify AGN and to trace central AGN activities in AGN. Then, considering the mean value of each 
distribution shown in Fig.~\ref{d1p}, properties of $R_{sii}$ are checked in AGN with each parameter larger than and smaller 
than the mean value. Here, the shown $L_{O3}$ are reddening corrected values through the measured Balmer decrements (flux 
ratio of narrow H$\alpha$ to narrow H$\beta$), after accepted the intrinsic Balmer decrement to be 3.1. And in the 
following subsections, there are no further discussions of reddening on our final results.

	Considering distributions of redshift, the estimated mean values of $\log(R_{sii})$ are about 0.041$\pm$0.003\ and 
0.048$\pm$0.004\ in the 2752 low redshift Type-1 AGN with $z<0.16$ and in the 2715 high redshift Type-1 AGN with $z>0.16$ 
in the main sample of the 5467 Type-1 AGN with $0.4~<~R_{sii}~<~1.5$, respectively. Estimated mean values of $\log(R_{sii})$ 
are about 0.073$\pm$0.003 and 0.068$\pm$0.003\ in the 4434 low redshift Type-2 AGN with $z<0.105$ and in the 3955 high 
redshift Type-2 AGN with $z>0.105$ in the main sample of the 8389 Type-2 AGN with $0.4~<~R_{sii}~<~1.5$, respectively. 
Therefore, considering different mean $\log(R_{sii})$ in different redshift ranges, there are accepted effects of different 
distributions of redshift on properties of distributions of calculated $R_{sii}$ in Type-1 AGN and in Type-2 AGN.

	Considering distributions of O3HB, the estimated mean values of $\log(R_{sii})$ are about 0.059$\pm$0.003\ and 
0.032$\pm$0.003\ in the 2544 Type-1 AGN with $\log(O3HB)$ smaller than 0.98 and in the 2923 Type-1 AGN with $\log(O3HB)$ 
larger than 0.98\ in the main sample of the 5467 Type-1 AGN with $0.4~<~R_{sii}~<~1.5$, respectively. The estimated 
mean values of $\log(R_{sii})$ are about 0.079$\pm$0.004\ and 0.062$\pm$0.003\ in the 4228 Type-2 AGN with $\log(O3HB)$ 
smaller than 0.74 and in the 4161 Type-2 AGN with $\log(O3HB)$ larger than 0.74\ in the main sample of the 8389 Type-2 
AGN with $0.4~<~R_{sii}~<~1.5$, respectively. Therefore, considering different mean $\log(R_{sii})$ in different O3HB 
ranges, there are also accepted effects of different distributions of O3HB on properties of distributions of calculated 
$R_{sii}$.

	Considering distribution of N2HA, the estimated mean values of $\log(R_{sii})$ are about 0.055$\pm$0.003\ and 
0.035$\pm$0.003\ in the 2584 Type-1 AGN with $\log(N2HA)$ smaller than -0.17 and in the 2883 Type-1 AGN with $\log(N2HA)$ 
larger than -0.17\ in the main sample of the 5467 Type-1 AGN with $0.4~<~R_{sii}~<~1.5$, respectively. The estimated 
mean values of $\log(R_{sii})$ are about 0.079$\pm$0.003\ and 0.062$\pm$0.003\ in the 4529 Type-2 AGN with $\log(N2HA)$ 
smaller than -0.072 and in the 3860 Type-2 AGN with $\log(N2HA)$ larger than -0.072\ in the main sample of the 8389 
Type-2 AGN with $0.4~<~R_{sii}~<~1.5$, respectively. Therefore, considering different mean $\log(R_{sii})$ in different 
N2HA ranges, there are also accepted effects of different distributions of N2HA on properties of distributions of 
calculated $R_{sii}$.

	Considering distribution of $L_{O3}$ in unit of ${\rm erg/s}$, the estimated mean values of $\log(R_{sii})$ 
are about 0.044$\pm$0.003\ and 0.046$\pm$0.003\ in the 2604 Type-1 AGN with $\log(L_{O3})$ smaller than 41.94 and in the 
2863 Type-1 AGN with $\log(L_{O3})$ larger than 41.94\ in the main sample of the 5467 Type-1 AGN with $0.4~<~R_{sii}~<~1.5$, 
respectively. The estimated mean values of $\log(R_{sii})$ are about 0.076$\pm$0.004\ and 0.066$\pm$0.003\ in the 3864 
Type-2 AGN with $\log(L_{O3})$ smaller than 41.32 and in the 4525 Type-2 AGN with $\log(L_{O3})$ larger than 41.32\ in 
the main sample of the 8389 Type-2 AGN with $0.4~<~R_{sii}~<~1.5$, respectively. Therefore, considering different mean 
$\log(R_{sii})$ in different $L_{O3}$ ranges, especially in Type-2 AGN, there are also accepted effects of different 
distributions of $L_{O3}$ on properties of distributions of calculated $R_{sii}$.


        Considering distribution of SN, the estimated mean values of $\log(R_{sii})$ are about 0.044$\pm$0.002\ and 
0.045$\pm$0.002\ in the 2650 Type-1 AGN with $\log(SN)$ smaller than 1.15 and in the 2817 Type-1 AGN with $\log(SN)$ 
larger than 1.15\ in the main sample of the 5467 Type-1 AGN with $0.4~<~R_{sii}~<~1.5$, respectively. The estimated 
mean values of $\log(R_{sii})$ are about 0.069$\pm$0.003\ and 0.073$\pm$0.003\ in the 4380 Type-2 AGN with $\log(SN)$ 
smaller than 1.21 and in the 4009 Type-2 AGN with $\log(SN)$ larger than 1.21\ in the main sample of the 8389 Type-2 AGN 
with $0.4~<~R_{sii}~<~1.5$, respectively. Meanwhile, as the shown results in subsection above, the calculated mean 
$R_{sii}$ in AGN in the main samples are different from the calculated $R_{sii}$ through emission line properties in 
mean spectra of the collected high quality AGN. Therefore, considering different mean $\log(R_{sii})$ in different SN 
ranges, there are also accepted effects of different distributions of SN on properties of distributions of calculated 
$R_{sii}$.

	Before proceeding further, one point is noted. Not similar as the physical quantities of $z$, O3HB, N2HA and 
$L_{O3}$, SN is a parameter related to spectra quality. Why effects of different SN are considered? Actually, there 
is negative dependence of SN on redshift in AGN. The Spearman Rank correlation coefficients are about -0.57 
($P_{null}<10^{-15}$) and -0.65 ($P_{null}<10^{-15}$) for the collected 5467 Type-1 AGN with $0.4~<~R_{sii}~<~1.5$ and 
for the 8389 Type-2 AGN with $0.4~<~R_{sii}~<~1.5$, respectively. Here, we do not show the dependence of SN on redshift 
in plots. However, considering the effects of different redshifts on $R_{sii}$ comparisons between Type-1 AGN and Type-2 
AGN, it is consequent to consider effects of different distributions of SN.

	Due to discussions above, it is necessary and interesting to check effects of different distributions of redshift, 
O3HB, N2HA, $L_{O3}$ and SN on the results in Fig~\ref{dis}. The convenient way is to create one subsample of Type-1 AGN 
which have the same distributions of redshift, O3HB, N2HA, $L_{O3}$ and SN as those of a subsample of Type-2 AGN. Based 
on the distributions of $z$, O3HB, N2HA, $L_{O3}$ and SN of the AGN in the main samples shown in Fig.~\ref{d1p} (5467 
Type-1 AGN with $0.4~<~R_{sii}~<~1.5$ and 8389 Type-2 AGN with $0.4~<~R_{sii}~<~1.5$), it is easy to create a subsample 
of Type-2 AGN having the same distributions of $z$, O3HB, N2HA, $L_{O3}$ and $SN$ as those of the Type-1 AGN in the 
subsample, through finding minimum parameter distance $D_p~<~D_{cri}$ calculated as
\begin{equation}
\begin{split}
D_{p,~i}~&=~D_{z,~i}~+~D_{O3HB,~i}~+~D_{N2HA,~i}~+~D_{L_{O3},~i}~+~D_{SN,~i} \\
	&=~(\frac{z_{1, i} - z_2}{sca_z})^2~+~(\frac{\log(O3HB_{1,i}) - \log(O3HB_{2})}{sca_{O3HB}})^2\\
	&\ \ ~+~(\frac{\log(N2HA_{1,i}) - \log(N2HA_{2})}{sca_{N2HA}})^2\\
	&\ \ ~+~(\frac{\log(L_{O3,~1,~i}) - \log(L_{O3,~2})}{sca_{L_{O3}}})^2 \\
	&\ \ ~+~(\frac{\log(SN_{1,~i}) - \log(SN_{2})}{sca_{SN}})^2 \ \ \ \ \ for\ i=1,\dots, N_1
\end{split}
\end{equation}
where $z_{1,~i}$, $O3HB_{1,~i}$, $N2HA_{1,~i}$, $L_{O3,~1,~i}$ and $SN_{1,~i}$ mean parameters of the $i$th Type-1 AGN 
in the main sample with $0.4~<~R_{sii}~<~1.5$ ($N_1~=~5467$), $z_2$, $O3HB_{2}$, $N2HA_{2}$, $L_{O3,~2}$ and $SN_2$ 
mean parameters of $N_2~=~8389$ ($N_2~>~N_1$) Type-2 AGN in the main sample with $0.4~<~R_{sii}~<~1.5$, $sca_z$, 
$sca_{O3HB}$, $sca_{N2HA}$, $sca_{L_{O3}}$ and $sca_{SN}$ are scale factors leading to $D_z$, $D_{O3HB}$, $D_{N2HA}$ 
and $D_{L_{O3}}$ not much different in quantity, and $D_{cri}$ means a critical value to prevent high $D_p$ leading 
to much different distributions of $z$, O3HB, N2HA, $L_{O3}$ and SN between the created final two subsamples. Then, 
based on $sca_z~\sim~0.002$ and $sca_{O3HB}~\sim~0.01$, $sca_{N2HA}~\sim~0.007$ and $sca_{L_{O3}}~\sim~0.02$ and 
$sca_{SN}~\sim~0.0065$ and $D_{cri}~\sim~60$, one subsample of 548 Type-1 AGN and one subsample of 548 Type-2 AGN are 
created, which have the same distributions of $z$, O3HB, N2HA, $L_{O3}$ and $SN$ with significance levels higher than 
99\% through the two-sided Kolmogorov-Smirnov statistic technique. Certainly, each object in the main samples is 
selected once into the two subsamples. The distributions of $z$, O3HB, N2HA, $L_{O3}$ and SN for the AGN in the 
subsamples are shown in Fig.~\ref{d2p}.

	Simple descriptions are given as follows to determine the scale factors $D_z$, $D_{O3HB}$, $D_{N2HA}$ and 
$D_{L_{O3}}$ and the critical $D_{cri}$ by three steps. First, starting values of the scale factors are set to be 
the differences between the mean redshift, between the mean $\log(O3HB)$, between the mean $\log(N2HA)$, between the 
mean $\log(L_{O3})$ and between the mean $\log(SN)$ of the 5467 Type-1 AGN and the 8389 Type-2 AGN in the main samples 
with $0.4~<~R_{sii}~<~1.5$: $sca_z=0.05$, $sca_{O3HB}=0.25$, $sca_{N2HA}=0.1$, $sca_{L_{O3}}=0.65$, $sca_{SN}=0.06$. 
And the starting value of $D_{cri}=260$ is the mean value of $D_{p,~0}$ determined by the starting values of the scale 
factors. And then, based on the Equation (6), two subsamples are created. Second, for the created two subsamples, the 
two-sided Kolmogorov-Smirnov statistic technique is applied to check whether the two subsamples have the same 
distributions of $z$, O3HB, N2HA, $L_{O3}$ and $SN$ with significance levels higher than 99\%. If the created two 
subsamples have different distributions of $z$ and/or O3HB and/or N2HA and/or $L_{O3}$ and/or $SN$ (statistical 
significance level smaller than 99\%), then smaller values should be re-assigned to the corresponding scale factors 
and $D_{cri}$. Based on the re-given $D_z$, $D_{O3HB}$, $D_{N2HA}$ and $D_{L_{O3}}$ and $D_{cri}$, two new subsamples 
are created, and then to check whether the two new subsamples have the same distributions of $z$, O3HB, N2HA, $L_{O3}$ 
and $SN$ with significance levels higher than 99\%. Third, repeating the second step, until the created two subsamples 
have the same distributions of $z$, O3HB, N2HA, $L_{O3}$ and $SN$ with significance levels higher than 99\%. The two 
subsamples of 548 Type-1 AGN and 548 Type-2 AGN in the manuscript are created after 15 attempts. And the basic 
parameters of redshift, O3HB, N2HA, SN, $L_{O3}$ and $R_{sii}$ are listed in Table~1 and Table~2.

	Moreover, in order to further confirm the two created subsamples having intrinsically same physical properties 
of $z$, O3HB, N2HA, $L_{O3}$ and $SN$ between Type-1 AGN and Type-2 AGN, it is necessary to check whether the collected 
Type-1 AGN and Type-2 AGN with one fixed parameter have the same distributions of the other four parameters. Here, 
Type-2 AGN and Type-1 AGN are collected with absolute value of one of the five parameters minus its mean 
value\footnote{To select different point from the mean value can lead to the same results} smaller than 5\%\footnote{The 
critical value 5\% can lead about 100 Type-1 AGN and about 100 Type-2 AGN to be collected, leading to much clearer 
histogram distributions of the parameters.} of the total range of the parameter. Then, the two-sided Kolmogorov-Smirnov 
statistic technique is applied to checked whether the collected Type-1 AGN and Type-2 AGN from the subsamples having 
the same distributions of the other four parameters. The results are shown in Fig.~\ref{d3s}. It is apparent that the 
collected Type-1 AGN and Type-2 AGN with one given parameter have the same distributions of the other four parameters 
with significance level higher than 75\% (actually most of the cases have the significance levels higher than 90\%). 
Therefore, the collected 548 Type-1 AGN and 548 Type-2 AGN in the subsamples can be well and efficiently applied to check 
different $R_{sii}$ (to simply trace properties of $n_e$) properties between Type-1 AGN and Type-2 AGN, after 
considerations of necessary effects.

\begin{figure}
\centering\includegraphics[width = 8cm,height=5.5cm]{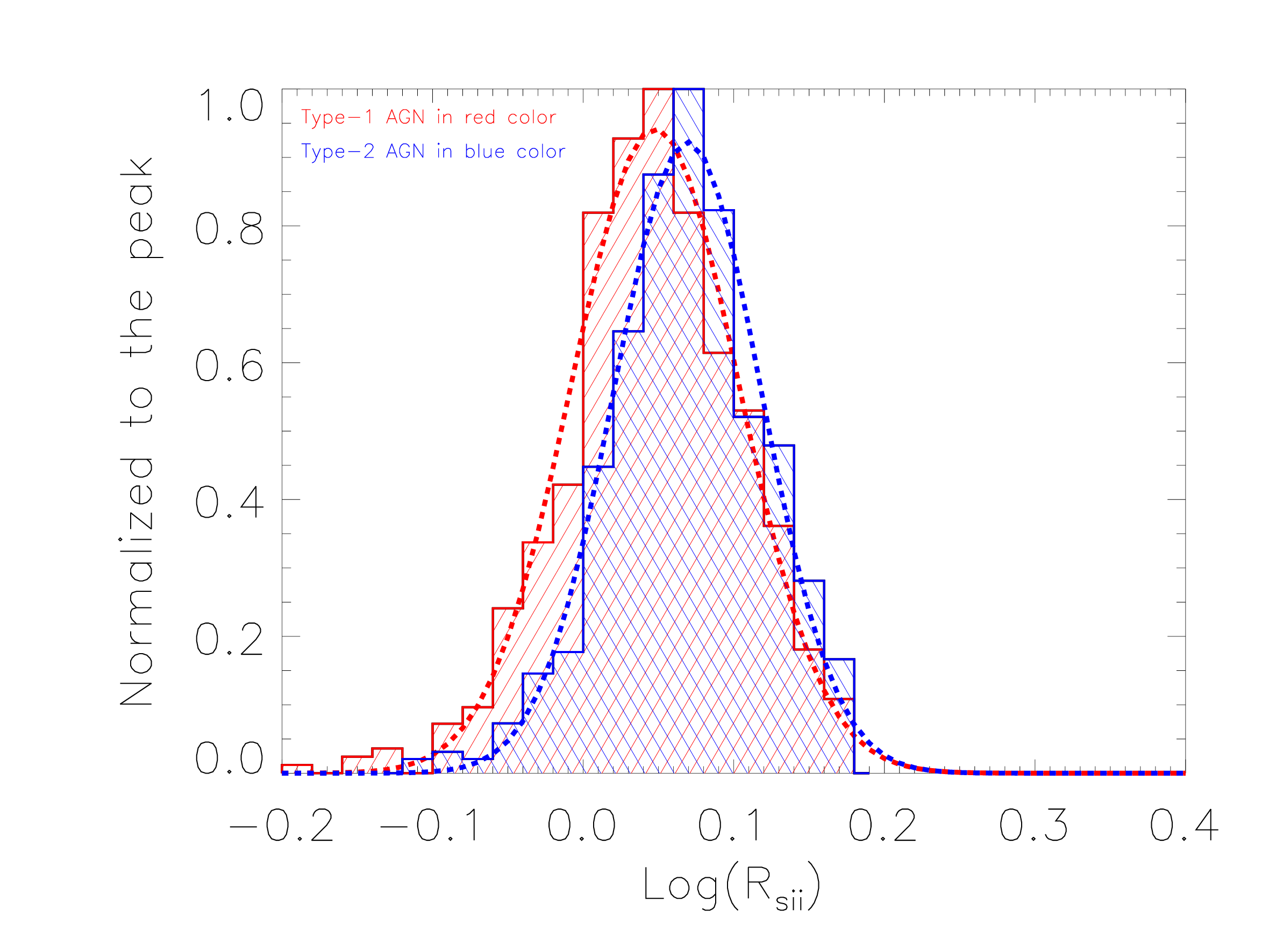}
\caption{Similar as Fig.~\ref{dis}, but for the 548 Type-1 AGN and the 548 Type-2 AGN in the subsamples, which have the 
same distributions of redshift, O3HB, N2HA, $L_{O3}$ and SN. The symbols and line styles have the same meanings as 
those in Fig.~\ref{dis}.}
\label{d2s}
\end{figure}

	Based on the subsamples of the 548 Type-1 AGN and the 548 Type-2 AGN, $R_{sii}$ distributions are shown in 
Fig.~\ref{d2s}, with mean $\log(R_{sii})$ about 0.042$\pm$0.005 and 0.072$\pm$0.005 of the Type-1 AGN and the Type-2 AGN, 
respectively, with uncertainties determined by the bootstrap method within 1000 loops. The new mean $R_{sii}$ can lead 
the corresponding mean $n_e$ in units of ${\rm cm^{-3}}$ to be estimated as 326$\pm$7 and 225$\pm$8 of the 548 Type-1 
AGN and the 548 Type-2 AGN in the subsamples. Therefore, after considering the necessary effects of different distributions 
of redshift, O3HB, N2HA, $L_{O3}$ and SN, Type-1 AGN have higher $n_e$ in NLRs than Type-2 AGN. 

	Furthermore, the well-known Students T-statistic technique is applied to confirm that the mean values of 
$\log(R_{sii})$ of the 548 Type-1 AGN and the 548 Type-2 AGN in the subsamples shown in Fig.~\ref{d2s} are significantly 
different with confidence level about $2.4\times10^{-10}$ (higher than 5$\sigma$). And the two-sided Kolmogorov-Smirnov 
statistic technique indicates that the Type-1 AGN and the Type-2 AGN obey the same distributions of $\log(R_{sii})$ 
with significance level about $8.4\times10^{-11}$ (higher than 5$\sigma$). Therefore, before giving clear effects 
of electron temperature on measurements of electron densities (which will be well discussed in the subsection 3.6), 
Type-1 AGN have apparently higher electron densities $n_e$ (only related to smaller $R_{sii}$) in NLRs than the Type-2 AGN, 
with confidence level higher than 5$\sigma$, against the expected results by the Unified model of AGN.

\subsection{Stronger AGN activities in Type-1 AGN?}

	Based on the higher $n_e$ in NLRs in Type-2 AGN than in the HII galaxies as the shown results in Fig.~\ref{dis}, 
AGN activities can be well applied to explain the higher $n_e$ in the Type-2 AGN than in HII galaxies, due to probable 
injecting electrons into NLRs through the galactic-scale outflows expected by the AGN feedback which plays key roles in 
galaxy evolution leading to tight connections between AGN and host galaxies as discussed in \citet{ref1, ref2, kh13, 
ref4, ref5, ref6, ref7}. If the AGN feedback expected outflows can lead to the higher $n_e$ in Type-2 AGN than in HII 
galaxies, the stronger outflows could be also well applied to explain the higher $n_e$ in NLRs in Type-1 AGN than in 
Type-2 AGN. More recently, \citet{kg18} have shown that there are statistical higher electron densities in NLRs in 
outflowing Seyfert galaxies than in non-outflowing Seyfert galaxies. Therefore, it is interesting to consider effects 
of outflows on our results.

	As discussed in \citet{cm14, ff17}, the kinetic powers of outflows are tightly scaled with AGN bolometric 
luminosity, indicating stronger outflows in AGN with strong [O~{\sc iii}] line luminosity. Similar results on the 
dependence of shifted velocities of [O~{\sc iii}] lines on continuum luminosity can be found in \citet{zh21a}. However, 
even the Type-2 AGN and the Type-1 AGN have the same properties of the [O~{\sc iii}] line luminosity, the higher 
$n_e$ can be also confirmed in Type-1 AGN as the results shown in Fig.~\ref{d2s}. Meanwhile, considering the fitting 
results to the emission lines around H$\alpha$ in the mean spectra of high quality Type-1 AGN and high quality Type-2 
AGN in Fig.~\ref{msp}, the [S~{\sc ii}] doublets have symmetric line profiles in Type-1 AGN and in Type-2 AGN, because 
the [S~{\sc ii}] doublets can be well described by two Gaussian components. If there were apparent effects of expected 
strong outflows on [S~{\sc ii}] doublets, there should be double-peaked features and/or asymmetric line profiles as 
shown in \citet{kg18}. Therefore, the symmetric line profiles of [S~{\sc ii}] doublets support that there are no 
apparent different properties of outflows in current stages in the Type-1 AGN and in the Type-2 AGN, and it is not necessary 
to consider effects of asymmetric wings in [S~{\sc ii}] doublets on our final results. Therefore, rather than the 
present injecting electrons into NLRs through galactic-scale outflows, longer durations of AGN activities triggering 
outflows in Type-1 AGN could be naturally applied to explain the higher $n_e$ in NLRs in Type-1 AGN.

	Either the higher $n_e$ in NLRs in Type-1 AGN or the expected long durations of AGN activities triggering 
outflows in Type-1 AGN are against the expected results by the commonly and widely accepted Unified model of AGN.

\subsection{Stronger star-forming contributions to NLRs in Type-2 AGN?}

	The main objective of the manuscript is to check the Unified Model of AGN through comparisons of electron 
densities in NLRs between Type-1 AGN and Type-2 AGN. Under the framework of the Unified Model of AGN, based on the 
same distributions of redshift between the 548 Type-1 AGN and the 548 Type-2 AGN in the subsamples, there are the 
same expected evolutionary histories between the Type-1 AGN and the Type-2 AGN in the subsamples, indicating the 
same host galaxy properties (including expected similar contributions of star-forming) between the 548 Type-1 AGN 
and the 548 Type-2 AGN in the subsamples.

	However, if there were more contributions from HII regions in Type-2 AGN than in Type-1 AGN, lower electron 
densities in NLRs would be expected in Type-2 AGN, due to lower electron densities in HII regions, as the shown 
results for HII galaxies in Fig.~\ref{dis}. However, the assumption that more star-forming contributions in Type-2 
AGN than in Type-1 AGN is against what have been expected by the Unified Model of AGN, to support our main final 
conclusion that the manuscript provide interesting clues to challenge the Unified Model of AGN.

\subsection{Further Discussions}

	In the discussed results above, effects of aperture sizes on the measured $R_{sii}$ are not considered. 
Actually, Type-2 AGN with lower redshift than 0.1 should have their emission regions of [S~{\sc ii}] doublet partly 
covered in the SDSS fiber spectra. However, if to consider the 140 Type-1 AGN and the 139 Type-2 AGN with redshift 
larger than 0.15 (corresponding fiber distance about 5200pc large enough to totally cover the NLRs of AGN with 
$L_{O3}\sim10^{41}{\rm erg\cdot~s^{-1}}$) in the subsamples, the mean $\log(R_{sii})$ are about 0.042$\pm$0.003 
and 0.072$\pm$0.004\ in the Type-1 AGN and in the Type-2 AGN, also leading to higher $n_e$ in NLRs of Type-1 AGN 
than Type-2 AGN. Here, the distances $R_{NLRs}$ of NLRs to central BHs in AGN are simply determined by the empirical 
relation between $R_{NLRs}$ and [O~{\sc iii}] line luminosity, as well discussed in \citet{liu13, ha13, ha14, fk18, 
dz18}. Therefore, effects of aperture sizes have few effects on the final results.


\begin{figure}
\centering\includegraphics[width = 8cm,height=6cm]{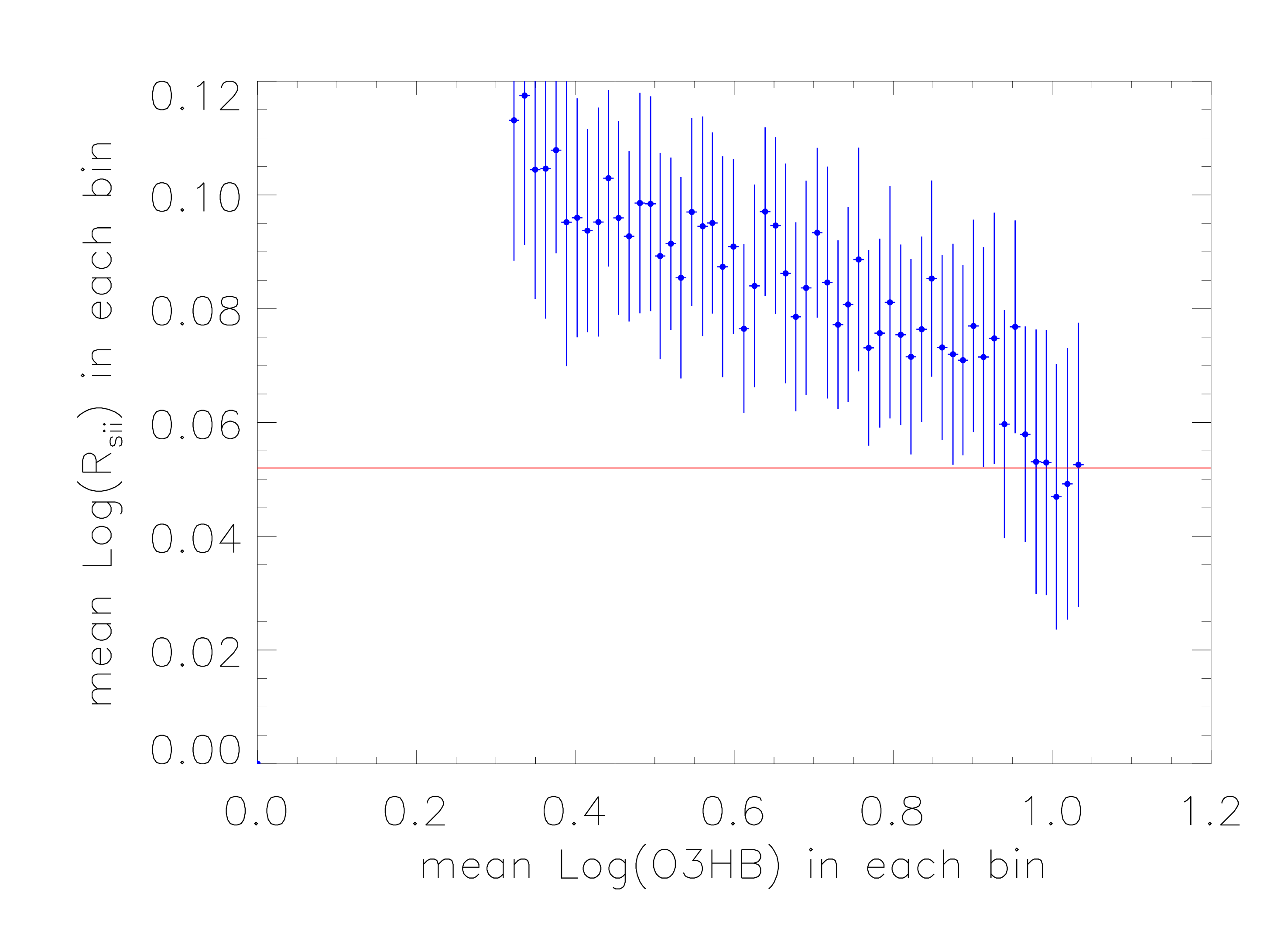}
\caption{On the dependence of mean $\log(R_{sii})$ on mean $\log(O3HB)$ for the main sample of the 8725 Type-2 AGN 
(excluding the Type-2 LINERs and composite galaxies) divided into 55 bins (at least 50 objects included in each bin) 
with equal width of $\log(O3HB)$. Horizontal red line marks the position $\log(R_{sii})=0.052$.
}
\label{ro3}
\end{figure}

	Moreover, as described in Section 2, we can totally confirm that Type-1 AGN cannot be mis-collected into HII 
galaxy sample or into Type-2 AGN sample, because apparent broad H$\alpha$ in Type-1 AGN but no broad H$\alpha$ in HII 
galaxies nor in Type-2 AGN, however, we cannot give the totally confirmed conclusion that there are no Type-2 AGN 
mis-collected into the HII galaxy sample. Therefore, effects are simply discussed on our final results, if some Type-2 
AGN were mis-collected into HII galaxies (or some HII galaxies mis-collected into Type-2 AGN). For the 8725 Type-2 AGN 
in the main sample, Fig.~\ref{ro3} shows the dependence of mean $\log(R_{sii})$ on mean $\log(O3HB)$ for the Type-2 AGN 
divided into 55 bins (at least 50 objects included in each bin) with equal width of $\log(O3HB)$. In Fig.~\ref{ro3}, 
uncertainty of each mean $\log(R_{sii})$ is calculated by the bootstrap method within 1000 loops. It is clear that in 
order to detect mean  $\log(R_{sii})$ to be about 0.042 (the mean $\log(R_{sii})$ for the Type-1 AGN in the main sample) 
in Type-2 AGN, the Type-2 AGN with $\log(O3HB)$ less than 1 should be the objects actually identified as HII galaxies, 
leading to the totally unreasonable results that about 95\% Type-2 AGN in the main sample were HII galaxies. Therefore, 
mis-collected HII galaxies into Type-2 AGN sample can not be applied to explain the different $n_e$ in NLRs between 
Type-1 AGN and Type-2 AGN. Meanwhile, considering the case that some Type-2 AGN were mis-collected into the HII galaxy 
sample, in order to detect mean $\log(R_{sii})$ to be about 0.042\ in Type-2 AGN, the HII galaxies with $\log(R_{sii})$ 
smaller than 0.042 should be the objects actually identified as Type-2 AGN. Among the HII galaxies in the main sample, 
there are 1269 HII galaxies with $\log(R_{sii})$ smaller than 0.052, even considering all the 1269 HII galaxies as 
Type-2 AGN, the mean $\log(R_{sii})$ is about 0.078$\pm$0.005\ in Type-2 AGN, re-confirming the higher electron density 
$n_e$ in NLRs in Type-1 AGN than in Type-2 AGN. 

\begin{figure*}
\centering\includegraphics[width = 18cm,height=10cm]{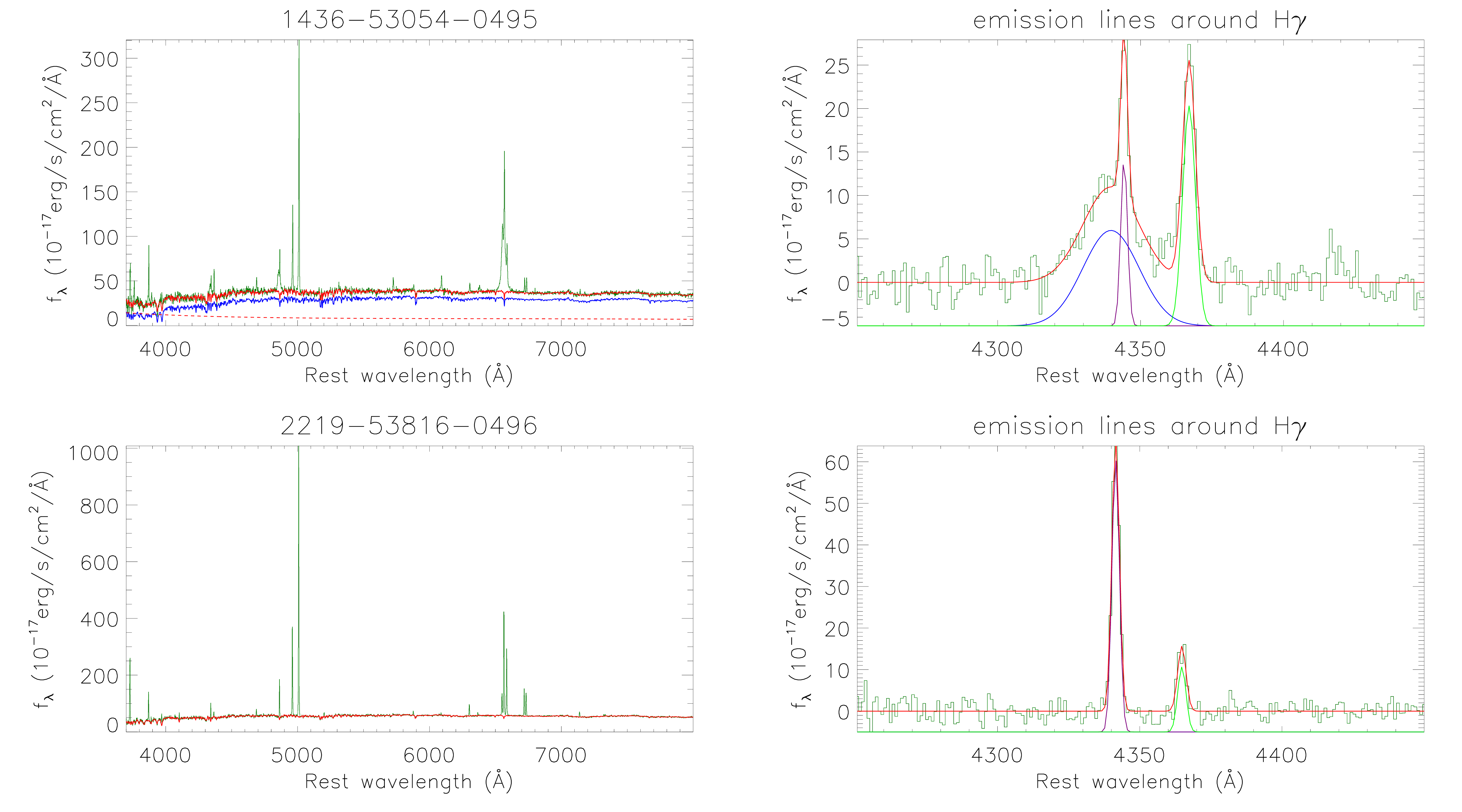}
\caption{Left panels show the SDSS spectra (in dark green) of the Type-1 AGN 1436-53054-0495 (top panel) and the Type-2 
AGN 2219-53816-0496 (bottom panel) in the subsamples, and the SSP method determined best descriptions (in red). In top 
left panel, solid blue line shows the determined host galaxy contributions and dashed red line shows the determined AGN 
continuum emissions. Right panels show the best descriptions (in red) to the emission lines around H$\gamma$ in the line 
spectrum (in dark green). In bottom region of each right panel, solid purple line and solid green line show the 
determined narrow H$\gamma$ and [O~{\sc iii}]$\lambda4364$\AA. And in bottom region of top right panel, solid blue line 
shows the determined broad H$\gamma$. Title of each left panel marks the information of PLATE-MJD-FIBERID of the SDSS spectrum.}
\label{pca}
\end{figure*}

\begin{figure*}
\centering\includegraphics[width = 18cm,height=10cm]{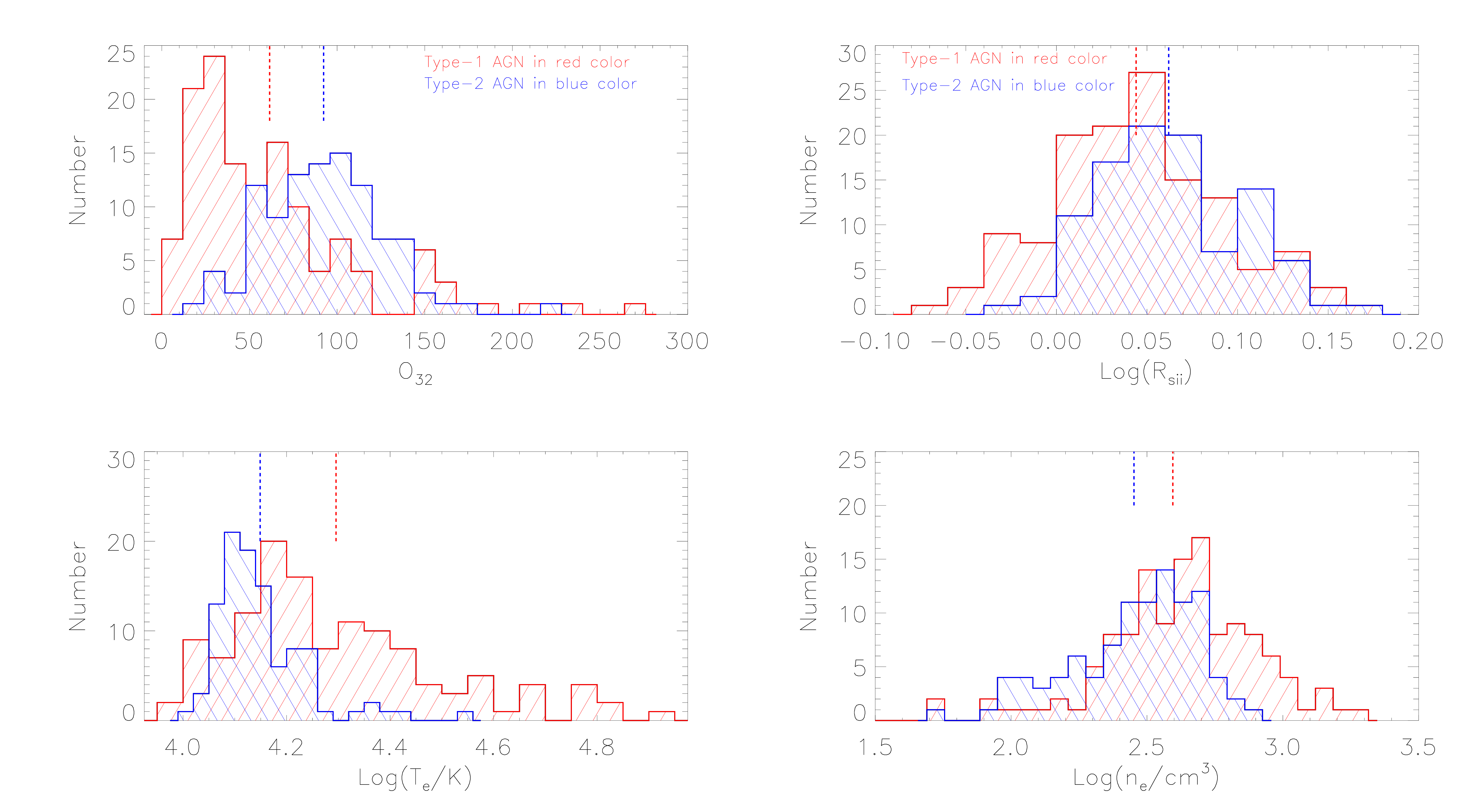}
\caption{Distributions of $O_{32}$ (top left panel), $\log(R_{sii})$ (top right panel), $\log(T_e/K)$ (bottom left panel) 
and improved electron density $\log(n_e/cm^3)$ (bottom right panel) of the 133 Type-1 AGN (in red color) and the 101 
Type-2 AGN (in blue color) which have apparent [O~{\sc iii}]$\lambda$4364\AA. In top region of each panel, vertical dashed 
lines in red and in blue mark the positions corresponding to the mean values of the Type-1 AGN and of the Type-2 AGN, respectively.}
\label{O4}
\end{figure*}

\begin{figure*}
\centering\includegraphics[width = 18cm,height=6cm]{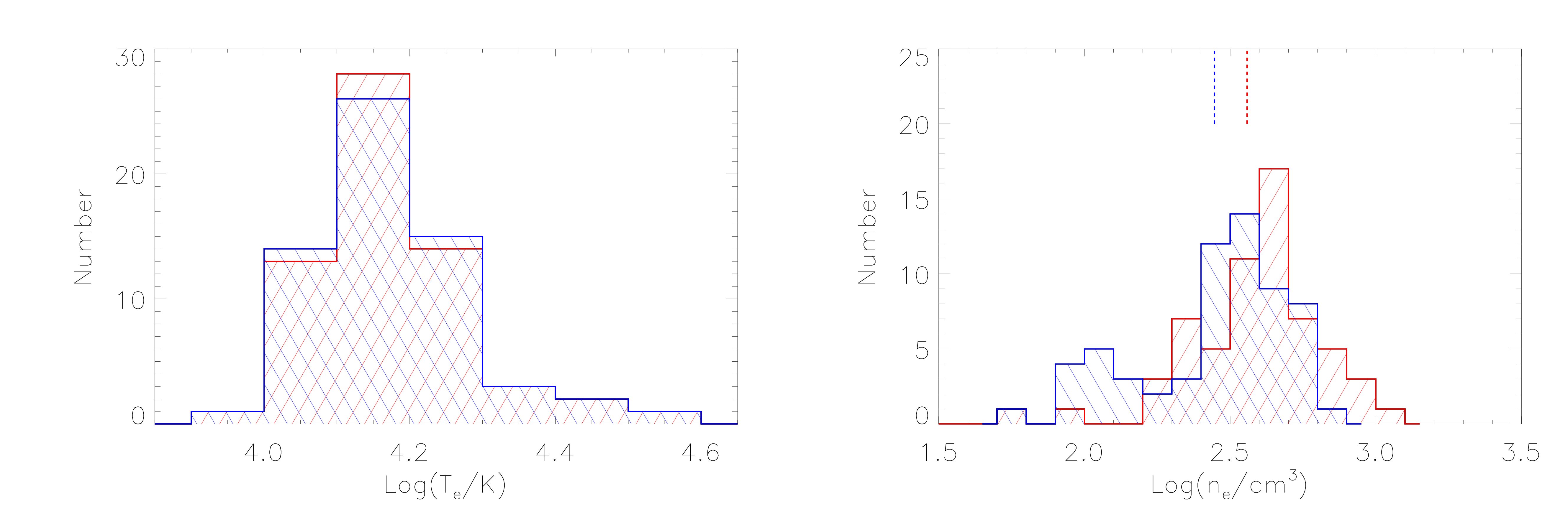}
\caption{The same $\log(T_e)$ distributions (left panel) and the $\log(n_e)$ distributions (right panel) of the 
62 Type-1 AGN and the 62 Type-2 AGN in the new created subsamples, In each panel, histogram filled by red lines shows the 
results for the 62 Type-1 AGN, and histogram filled by blue lines shows the results for the 62 Type-2 AGN. In right panel, 
vertical dashed red line and dashed blue line mark positions for mean values of $\log(n_e)$ of the 62 Type-1 AGN and the 
	62 Type-2 AGN, respectively.}
\label{ste}
\end{figure*}

\begin{figure*}
\centering\includegraphics[width = 18cm,height=8cm]{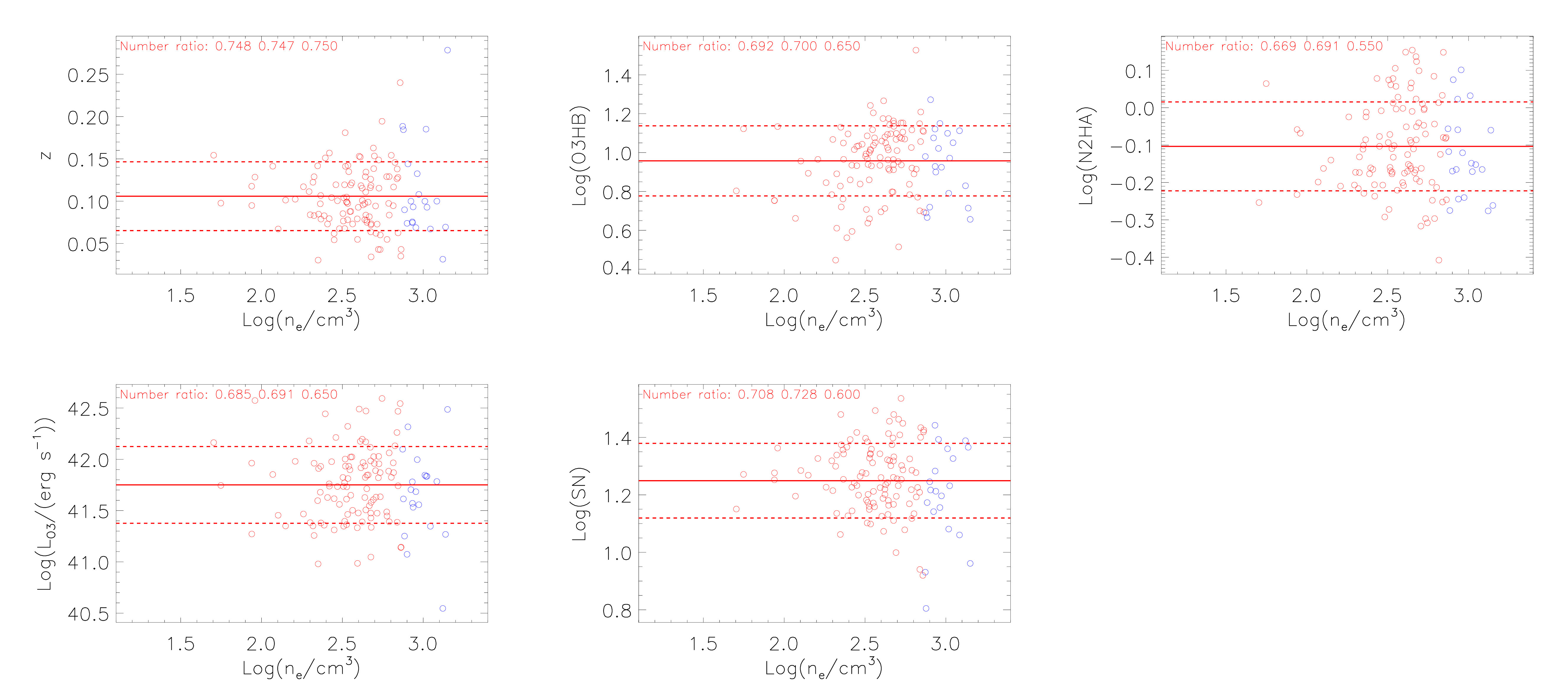}
\centering\includegraphics[width = 18cm,height=8cm]{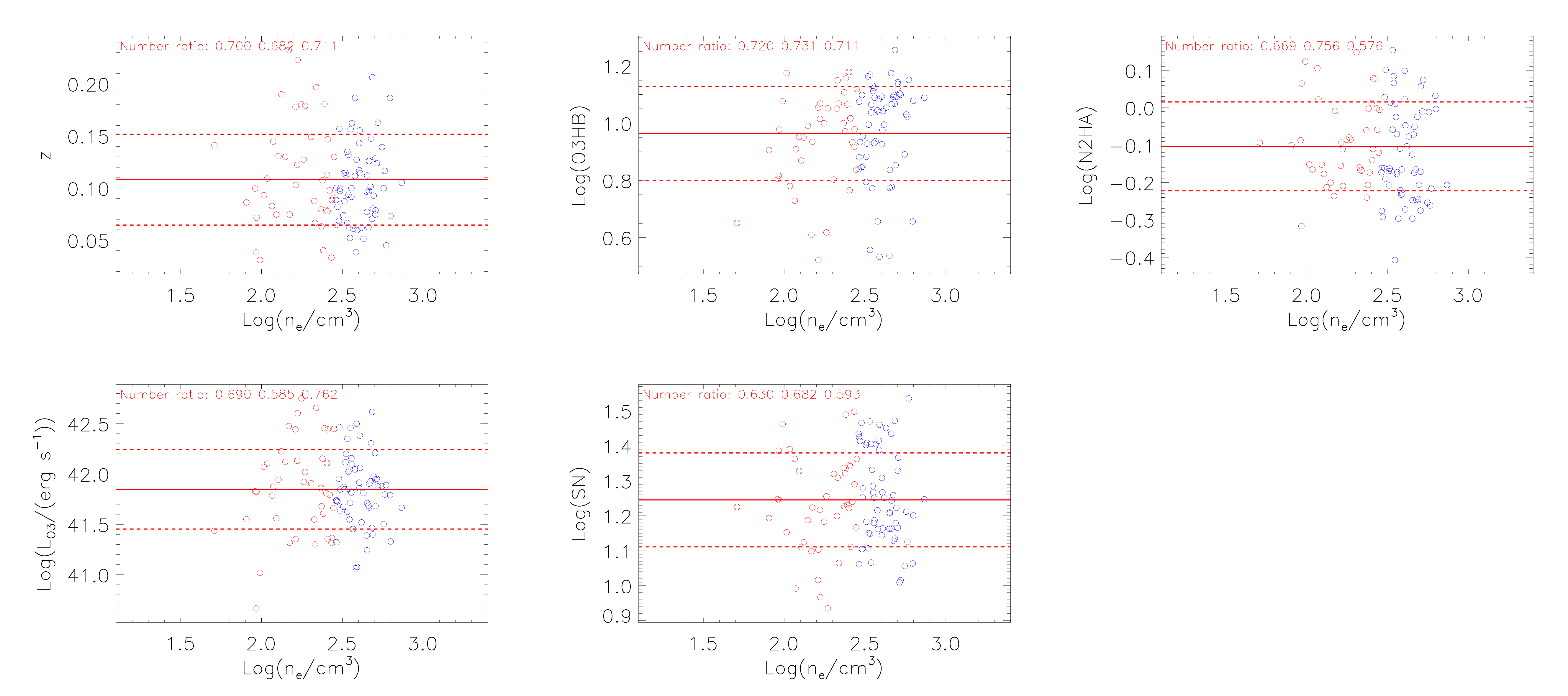}
\caption{On the correlations between $n_e$ and the parameters of $z$, O3HB, N2HA, $L_{O3}$ and SN for the Type-1 
AGN (panels in the first rows) and the Type-2 AGN (panels in the last rows) shown in bottom right panel of Fig~\ref{O4}. In 
each panel in the first two rows, open circles in red and in blue show the results for the Type-1 AGN with $\log(n_e)<2.87$ 
and the Type-1 AGN with $\log(n_e)>2.87$, respectively. Horizontal solid red line and horizontal dashed red lines show the 
mean value of the parameter shown in Y-axis and corresponding 1RMS scatter bands for all the Type-1 AGN. In each panel in 
the last two rows, open circles in red and in blue show the results for the Type-2 AGN with $\log(n_e)<2.45$ and the Type-2 
AGN with $\log(n_e)>2.45$, respectively. Horizontal solid red line and horizontal dashed red lines show the mean value of the 
parameter shown in Y-axis and corresponding 1RMS scatter bands for all the Type-2 AGN. In top region of each panel of the 
Figure, the three number ratios are marked for the ratio of AGN outsides of the 1RMS scatter bands to all the AGN, the ratio 
of AGN shown in blue outsides of the 1RMS scatter bands to all the AGN shown in blue, and the ratio of AGN shown in red 
outsides of the 1RMS scatter bands to all the AGN shown in red.}
\label{HL}
\end{figure*}

\begin{figure*}
\centering\includegraphics[width = 18cm,height=6cm]{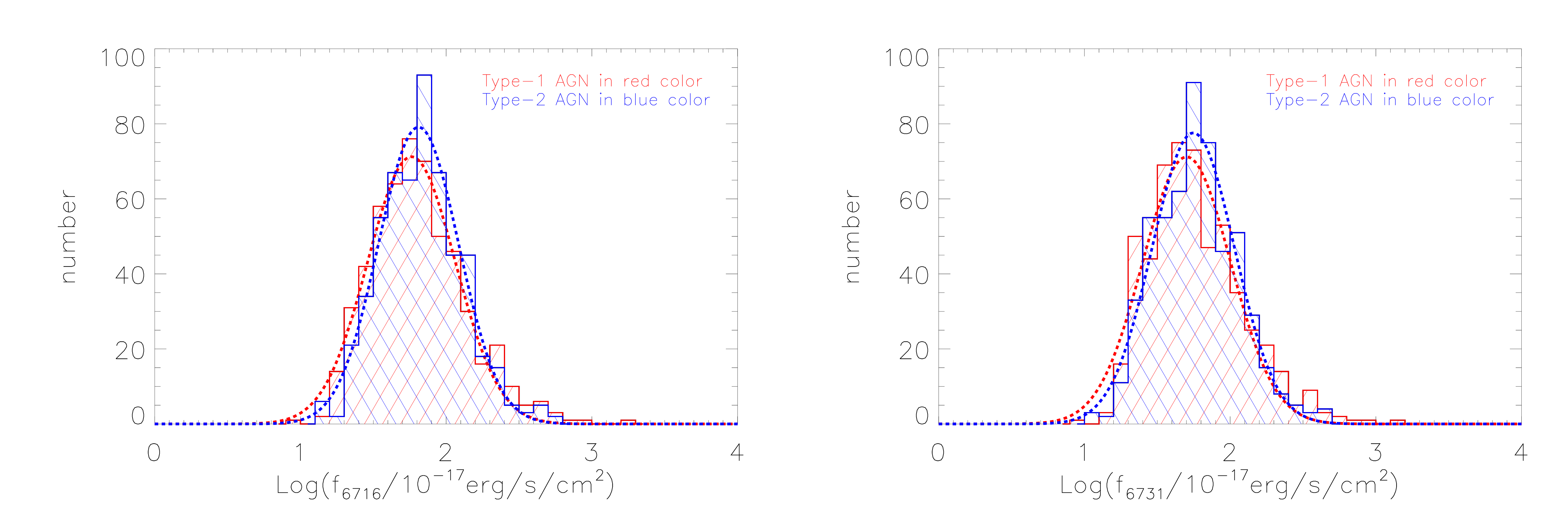}
\caption{Distributions of $f_{6716}$ (left panel) and $f_{6731}$ (right panel) in the 548 Type-1 AGN (in red color) and 
in the 548 Type-2 AGN (in blue color) in the subsamples. Thick dashed lines in red and in blue represent the corresponding 
best Gaussian profiles for the distributions of the 548 Type-1 AGN and the 548 Type-2 AGN in the subsamples, respectively.}
\label{fs22}
\end{figure*}

	Furthermore, as detailed discussions in \citet{os89, of06, kn19, fm20}, etc., there are apparent effects of 
electron temperature $T_e$ on estimating electron density $n_e$ by the parameter $\log(R_{sii})$, and the improved formula 
to estimate electron density can be described as (see Fig.~5.8 and corresponding discussions in \citet{of06}),
\begin{equation}
\frac{n_e}{\rm cm^{3}}\times(\frac{10^4K}{T_e})^{0.5}~\approxeq~\frac{627.1~\times~R_{sii}~-~909.17}{0.4315~-~R_{sii}}
\end{equation}
after considering effects of electron temperature $T_e$. Therefore, it is necessary to consider effects of $T_e$ on 
reported results on larger $n_e$ (smaller $\log(R_{sii})$) in Type-1 AGN. Electron temperatures $T_e$ can be well 
traced by flux ratio $O_{32}$ of the [O~{\sc iii}] lines
\begin{equation}
\begin{split}
&O_{32}=\frac{f_{\lambda4959}~+~f_{\lambda5007}}{f_{\lambda4364}}=
	\frac{7.9\times exp(\frac{3.29\times10^4K}{T_e})}{1+4.5\times10^{-4}n_e/T_e^{0.5}} \\
	&~~~~~~~~~~\sim7.9\times exp(\frac{3.29\times10^4}{T_e})\\
\end{split}
\end{equation}.

	For the 548 Type-1 AGN and the 548 Type-2 AGN in the subsamples, emission lines around [O~{\sc iii}]$\lambda4364$\AA~ 
within rest wavelength range from 4250\AA~ to 4450\AA~ are well measured by multiple-Gaussian functions, one narrow Gaussian 
function applied to describe narrow H$\gamma$, one narrow Gaussian function applied to describe narrow 
[O~{\sc iii}]$\lambda4364$\AA, two broad Gaussian functions applied to describe broad H$\gamma$ only in Type-1 AGN, 
after subtractions of host galaxy contributions (if there are) which have been determined above through the SSP method. 
Fig.~\ref{pca} shows one Type-1 AGN and one Type-2 AGN of which apparent [O~{\sc iii}]$\lambda4364$\AA~ are best described 
by multiple Gaussian functions. Then, through the criterion that the measured flux and second moment of [O~{\sc iii}]$\lambda4364$\AA~ 
at least 3 times larger than their corresponding uncertainties, there are 133 Type-1 AGN and 101 Type-2 AGN which have apparent 
[O~{\sc iii}]$\lambda4364$\AA. The results indicate that only a small part of AGN have apparent 
[O~{\sc iii}]$\lambda4364$\AA. That is the main reason why we do create our main samples (discussed in section 2) of AGN 
without considering properties of [O~{\sc iii}]$\lambda4364$\AA. Not similar as [O~{\sc iii}]$\lambda4959,5007$\AA~ doublet 
which are commonly clear and strong in AGN, [O~{\sc iii}]$\lambda4364$\AA~ emissions are commonly weak in AGN, leading to 
less number of AGN which have apparent [O~{\sc iii}]$\lambda4364$\AA~ emission features. If not only apparent H$\alpha$, 
H$\beta$, [O~{\sc iii}]$\lambda4959,5007$\AA, [N~{\sc ii}] and [S~{\sc ii}] but also apparent [O~{\sc iii}]$\lambda4364$\AA~ 
(line parameters are at least 5 times larger than their corresponding uncertainties) were considering to create new main 
samples, only about one seventh of the AGN in the main samples created in Section 2 were retained into new created main 
samples. Then, there should be only tens of AGN included in expected new subsamples which have the same distributions of 
$z$, O3HB, N2HA, $L_{O3}$ and $SN$, leading to not reliable discussions on results through the new created subsamples.

	Then, distributions of $O_{32}$ and $\log(R_{sii})$ are shown in top panels of Fig.~\ref{O4}, with mean values 
[$O_{32}$,~$\log(R_{sii})$] of [61.49$\pm$6.42,~0.044$\pm$0.006] in the 133 Type-1 AGN and of [92.27$\pm$4.12,~0.062$\pm$0.005] 
in the 101 Type-2 AGN, respectively. Based on properties of $O_{32}$, bottom left panel of Fig.~\ref{O4} shows distributions 
of $T_e$ which are also listed in Table~1 and Table~2, with mean values of $(1.95\pm0.14)\times10^4K$ in the 
133 Type-1 AGN and of $(1.41\pm0.07)\times10^4K$ in the 101 Type-2 AGN, respectively. Then, based on the calculated $T_e$ 
and $\log(R_{sii})$, bottom right panel of Fig.~\ref{O4} shows distributions of $n_e/cm^3$ after corrections of effects 
of $T_e$. The improved mean electron densities $n_e/{\rm cm^3}$ are about $394\pm36$ and $283\pm23$ in the 133 Type-1 AGN 
and in the 101 Type-2 AGN, respectively, re-leading to apparently large $n_e$ in NLRs in Type-1 AGN than in Type-2 AGN. 
Uncertainties of the mean values above are determined by the bootstrap method within 1000 loops. Furthermore, the two-sided 
Kolmogorov-Smirnov statistic technique is applied to determine that the 133 Type-1 AGN and the 101 Type-2 AGN obey the same 
distributions of $\log(R_{sii})$ with significance level only about $6\times10^{-5}$ (higher than 4$\sigma$). And the Students 
T-statistic technique is applied to confirm that the mean values of $n_e$ of the 133 Type-1 AGN and the 101 Type-2 AGN in 
the subsamples are significantly different with confidence level about $6.9\times10^{-7}$ (higher than 5$\sigma$). 
Moreover, as shown in bottom right panel of Fig.~\ref{O4}, it looks like there is a cut value $\log(n_e)\sim2.87$ 
for Type-2 AGN, it is also necessary to roughly check whether the cut value can lead to different results on $n_e$. Here, 
even accepted $log(n_e)\sim2.87$ (the maximum value for Type-2 AGN) as a cut value, the mean values of $log(n_e)$ are 
2.51$\pm$0.03 (323$\pm$21$cm^3$) and 2.45$\pm$0.02 (282$\pm$14$cm^3$) for the 107 Type-1 AGN with $log(n_e)<2.87$ and the 
101 Type-2 AGN with $\log(n_e)<2.87$, leading to higher $n_e$ in Type-1 AGN than in Type-2 AGN. Also, the two-sided 
Kolmogorov-Smirnov statistic technique can lead to probability only about $1.9\times10^{-2}$ to support similar $\log(n_e)$ 
distributions of the 107 Type-1 AGN with $\log(n_e)<2.87$ and the 101 Type-2 AGN with $\log(n_e)<2.87$, and the Students 
T-statistic technique can lead to probability only about $4.5\times10^{-3}$ to support similar mean values of $n_e$ of the 
107 Type-1 AGN with $\log(n_e)<2.87$ and the 101 Type-2 AGN with $\log(n_e)<2.87$. Therefore, effects of electron 
temperatures can be applied to re-confirm the larger electron densities in NLRs of Type-1 AGN than in Type-2 AGN.


	Moreover, based on $T_e$ distributions shown in Fig.~\ref{O4}, effects of different $T_e$ distributions can be 
checked by the following method, as what have been done in subsection 3.3. Through equation (6) with applications of only 
one parameter $\log(T_e)$ of the 133 Type-1 AGN and the 101 Type-2 AGN shown in Fig.~\ref{O4}, one subsample of 62 Type-1 
AGN and the other one subsample of 62 Type-2 AGN can be created with the same $\log(T_e)$ distributions. The two-sided 
Kolmogorov-Smirnov statistic technique can lead to probability 98.4\% to support the same $\log(T_e)$ distributions of 
AGN in the two subsamples. Left panel of Fig.~\ref{ste} shows the $\log(T_e)$ distributions of AGN in the two subsamples. 
Then, right panel of Fig.~\ref{ste} shows the $n_e$ distributions of AGN in the two subsamples. The mean values of $\log(n_e)$ 
are about 2.56$\pm$0.04 (363$\pm$33${\rm cm^{-3}}$) and 2.45$\pm$0.03 (282$\pm$19${\rm cm^{-3}}$) for the 62 Type-1 AGN 
and the 62 Type-2 AGN in the subsamples having the same $\log(T_e)$ distributions, leading to higher $n_e$ in the 62 Type-1 
AGN than in the Type-2 AGN in the new created subsamples. Uncertainties of the mean values above are determined by the bootstrap 
method within 1000 loops. Also, the two-sided Kolmogorov-Smirnov statistic technique can lead to probability only about 
$2.6\times10^{-2}$ to support similar $\log(n_e)$ distributions of the 62 Type-1 AGN and the 62 Type-2 AGN in the new 
created subsamples, and the Students T-statistic technique can lead to probability only about $3.1\times10^{-2}$ to support 
similar mean values of $\log(n_e)$ of the 62 Type-1 AGN and the 62 Type-2 AGN in the new created subsamples. Therefore, 
totally ignoring effects of different $T_e$ distributions shown in Fig.~\ref{O4}, higher electron density in NLRs in 
Type-1 AGN can be well confirmed, even there are less numbers of AGN in the new created subsamples.

	Furthermore, through the shown results in Fig.~\ref{O4}, it is interesting to consider whether the Type-1 AGN with 
higher electron densities $\log(n_e)>2.87$ have different physical properties from the other Type-1 AGN with $\log(n_e)<2.87$. 
Then, panels in the first two rows of Fig.~\ref{HL} shows correlations between $n_e$ and the parameters of $z$, SN, O3HB, N2HA 
and $L_{O3}$ of the Type-1 AGN shown in Fig.~\ref{O4}. The correlations have Spearman Rank correlation coefficients smaller 
than 0.2, thus rather than to determine linear fitting results (with determined parameters smaller than corresponding 
uncertainties by the FITEXY code), the mean values of $z$, SN, O3HB, N2HA and $L_{O3}$ and corresponding 1RMS scatter bands 
are shown in each panel of Fig.~\ref{HL}. It is clear that there are the similar number ratios (marked in each panel) of 
Type-1 AGN outside of the 1RMS scatter bands to all Type-1 AGN, of the Type-1 AGN with $\log(n_e)<2.87$ outside of the 1RMS 
scatter bands to all the Type-1 AGN with $\log(n_e)<2.87$, and of the Type-1 AGN with $\log(n_e)>2.87$ outside of the 1RMS 
scatter bands to all the Type-1 AGN with $\log(n_e)>2.87$. The similar number ratios strongly indicate that the Type-1 AGN 
with $\log(n_e)>2.87$ are not outliers among the reported Type-1 AGN. Meanwhile, the panels in the last two rows shows similar 
results for the Type-2 AGN shown in Fig.~\ref{O4} with accepted cut value $\log(n_e)\sim2.45$ (the mean value of $\log(n_e)$ 
of the Type-2 AGN). Similar results can be found that there are similar number ratios (marked in each panel) of Type-2 AGN 
outside of the 1RMS scatter bands to all Type-2 AGN, of the Type-2 AGN with $\log(n_e)<2.45$ outside of the 1RMS scatter 
bands to all the Type-2 AGN with $\log(n_e)<2.45$, and of the Type-2 AGN with $\log(n_e)>2.45$ outside of the 1RMS scatter 
bands to all the Type-2 AGN with $\log(n_e)>2.45$. Therefore, the results in Fig.~\ref{HL} strongly support that the selected 
AGN with different $n_e$ have similar physical properties as the other reported AGN in the manuscript.

	Furthermore as well discussed in \citet{fh84} with two zones with different electron densities in 
NGC 7213, if there were higher electron density regions closer to central BHs could be visible in Type-1 AGN but probably 
seriously obscured in Type-2 AGN, higher electron densities (smaller $R_{sii}$) could be well expected in Type-1 AGN. 
To put it simply, if the two-zone model was accepted, line fluxes ($f_{6716}$, $f_{6731}$) of each [S~{\sc ii}] line include
two components, one component ($f_{6716,H}$, $f_{6731,H}$) from higher electron density regions and the other one component
($f_{6716,L}$ and $f_{6731,L}$) from normal (or lower) electron density regions. Meanwhile, due to $f_{6716,H}$, $f_{6731,H}$ 
from higher electron density regions, we have
\begin{equation}
f_{6716,H}/f_{6731,H}~<~f_{6716,L}/f_{6731,L}
\end{equation}.
Then, based on serious dependence of flux ratios of [S~{\sc ii}] on electron densities, mean electron densities $n_e$ can be
simply determined by flux ratios of the two components as follows,
\begin{equation}
\begin{split}
&R_{sii} = \frac{f_{6716,H}+f_{6716,L}}{f_{6731,H}+f_{6731,L}} \\
&n_{e}\times(\frac{10^4K}{T_{e}})^0.5\sim\frac{627.1~\times~R_{sii}~-~909.17}{0.4315~-~R_{sii}}
\end{split}
\end{equation}.
For type-1 AGN, due to no obscurations, all the parameters from observed line fluxes can be well accepted intrinsic values.
However, for Type-2 AGN with orientation effects leading to higher electron density zones being seriously obscured, without 
contributions of $f_{6716,H}$ and $f_{6731,H}$, $R_{sii}$ in Type-2 AGN should be 
$R_{sii, T2} = \frac{f_{6716,L}}{f_{6731,L}}$. Considering $f_{6716,H}/f_{6731,H}~<~f_{6716,L}/f_{6731,L}$,
we will clearly have
\begin{equation}
R_{sii, T2}~>~R_{sii, T1}~=~\frac{f_{6716,H}+f_{6716,L}}{f_{6731,H}+f_{6731,L}}
\end{equation}
with $R_{sii, T1}$ as measurements of $R_{sii}$ in Type-1 AGN. 
That is why two zone model can be applied to explain higher electron densities in Type-1 AGN based on flux ratio $R_{sii}$ of
[S~{\sc ii}] emission lines. Meanwhile, after considering orientation effects expected obscurations on higher electron density
regions, flux intensities of [S~{\sc ii}] emission lines in Type-2 ($f_{6716,L}$ and $f_{6731,L}$) should be smaller than
the flux intensities ($f_{6716,L}+f_{6716,H}$ and $f_{6731,L}+f_{6731,H}$) in Type-1 AGN. Therefore, it is interesting and 
necessary to check effects of the probable higher electron density regions closer to central BHs on our final results. However, 
the mean [S~{\sc ii}]$\lambda6716$\AA~ ([S~{\sc ii}]$\lambda6731$\AA~) line intensities $\log(f_{6716}/{\rm 10^{-17}erg/s/cm^2})$ 
($\log(f_{6731}/{\rm 10^{-17}erg/s/cm^2})$) are about 1.812$\pm$0.013 (1.767$\pm$0.010) and 1.832$\pm$0.013 (1.765$\pm$0.010)\ 
in the 548 Type-1 AGN and the 548 Type-2 AGN in the subsamples respectively. The uncertainties are determined by the bootstrap 
method within 1000 loops. Distributions of $f_{6716}$ and $f_{6731}$ are shown in Fig.~\ref{fs22}. Slightly higher 
[S~{\sc ii}]$\lambda6716$\AA~ line intensities can be found in Type-2 AGN than in Type-1 AGN, and similar 
[S~{\sc ii}]$\lambda6731$\AA~ line intensities can be found between Type-1 AGN and Type-2 AGN. Because the Type-1 AGN and the 
Type-2 AGN in the subsamples have the same redshift distributions, [S~{\sc ii}] line luminosity distributions are not shown 
and discussed again. The results are against the expected higher [S~{\sc ii}] line intensities in Type-1 AGN, therefore, 
higher electron density regions visible in Type-1 AGN (or higher electron density regions partly obscured in Type-2 AGN) 
can not be applied to well explain the higher electron densities in Type-1 AGN than in Type-2 AGN. 

	Finally, simple discussions are given on probable asymmetric components in [S~{\sc ii}] doublet, although there 
are no clear clues to support apparent asymmetric components in [S~{\sc ii}] doublet in the mean spectra of high quality 
AGN. As is known, asymmetric components in [S~{\sc ii}] doublet could be tightly related to radial outflows. And galactic 
outflows could be tightly related to radio emissions, such as the more recent results in \citet{jm19}. Therefore, among 
the AGN in the main samples, radio properties are checked through the FIRST (Faint Images of the Radio Sky at Twenty-Centimeters) 
database \citep{bw95, hw15}, and to show whether are there quite different $\log(R_{sii})$ between AGN without radio 
emissions and AGN with radio emissions. Among the 6039 Type-1 AGN collected from SDSS DR12, there are 4432 Type-1 AGN 
(no-radio Type-1 AGN) covered by FIRST but with radio emission intensity to be zero, and 1607 Type-1 AGN (radio Type-1 AGN) 
covered by FIRST with radio emission intensity larger than zero. Mean $\log(R_{sii})$ are about 0.052$\pm$0.003 for the 
4432 no-radio Type-1 AGN and 0.049$\pm$0.003 for the 1607 radio Type-1 AGN, respectively. Meanwhile, among the 12999 
Type-2 AGN collected from SDSS DR12, there are 10431 Type-2 AGN (no-radio Type-2 AGN) covered by FIRST but with radio 
emission intensity to be zero, and 2568 Type-2 AGN (radio Type-2 AGN) covered by FIRST with radio emission intensity 
larger than zero. Mean $\log(R_{sii})$ are about 0.088$\pm$0.005 for the 10431 no-radio Type-2 AGN and 0.082$\pm$0.005 
for the 2568 radio Type-2 AGN. Therefore, even considering effects of asymmetric components related to radio emissions, it 
can be re-confirmed that Type-1 AGN has higher $n_e$ in NLRs than Type-2 AGN.

\section{Further Implications}

	If the higher electron density $n_e$ in NLRs in Type-1 AGN are intrinsically true, there could be some special 
type-1 AGN of which NLRs have electron densities high enough to be nearer to critical densities to forbidden emission 
lines, once strong injecting electrons can last long enough, leading to quite weak narrow forbidden emission lines. 
Therefore, in the near future, to detect so special Type-1 AGN without narrow forbidden emission lines is the main 
objective of one of our being prepared manuscripts.

\section{Summary and Conclusions}

	Finally, the main summary and conclusions are as follows.
\begin{itemize}
\item All the low redshift ($z~<~0.3$) Type-1 AGN and Type-2 AGN are collected from SDSS DR12. The 
	[S~{\sc ii}]$\lambda6716, 6731$\AA~ doublets are well measured. Based on the reliable 
	[S~{\sc ii}]$\lambda6716, 6731$\AA~ doublets, there are 6039 Type-1 AGN with reliable [S~{\sc ii}] doublets and 
	apparent broad H$\alpha$ emission lines, and 12999 Type-2 AGN with reliable [S~{\sc ii}] doublets but no broad 
	H$\alpha$ emission lines.
\item After considering the controversial conclusion on physical nature of Type-2 LINERs (at least part of Type-2 AGN 
	without AGN nature) and considering strong contributions from star-forming in composite galaxies, both Type-2 
	LINERs and composite galaxies are not considered in the final main sample of Type-2 AGN, lead the final main sample 
	of Type-2 AGN including 8725 Type-2 AGN.
\item Based on the reliable line flux ratio $R_{sii}$ of [S~{\sc ii}]$\lambda6716$\AA~ to [S~{\sc ii}]$\lambda6731$\AA, 
	higher electron densities $n_e$ in NLRs can be found in Type-1 AGN than in Type-2 AGN, and in Type-2 AGN than in 
	HII galaxies.
\item After considering necessary effects of redshift and central AGN activities on the distributions of $n_e$, two 
	subsamples of 548 Type-1 AGN and 548 Type-2 AGN are created to have the same distributions of $z$, O3HB, N2HA, 
	$L_{O3}$ and SN, still leading to higher $n_e$ in NLRs of Type-1 AGN than Type-2 AGN, with confidence level 
	higher than $5\sigma$.
\item Comparing the $n_e$ in NLRs between HII galaxies and Type-2 AGN, AGN activities related to central BH accreting power 
	should play key roles in higher electron densities in NLRs due to injecting electrons by AGN feedback expected 
	galactic-scale outflows.
\item Unfortunately, even Type-1 AGN and Type-2 AGN have the same properties of O3HB, N2HA and $L_{O3}$ at present times, 
	the AGN in the subsamples, higher $n_e$ in NLRs in Type-1 AGN can also be confirmed. Therefore, longer time 
	durations of AGN activities in Type-1 AGN should be preferred.
\item Considering lower electron densities in HII galaxies, stronger star-forming contributions to NLRs could be applied 
	to explain the lower electron densities in NLRs in Type-2 AGN than in Type-1 AGN, if not considering the Unified 
	Model expected similar host galaxy evolutionary histories for the AGN in the subsamples with the same distributions 
	of redshift.
\item After considering probable effects of asymmetric components in [S~{\sc ii}] doublets related to radio emissions, it 
	can be re-confirmed that Type-1 AGN without (with) radio emissions has higher $n_e$ in NLRs than Type-2 AGN without 
	(with) radio emissions.
\item After considering effects of electron temperatures traced by flux ratio of [O~{\sc iii}]$\lambda4364,4959,5007$\AA~ 
	emission lines on estimating electron densities in NLRs, more apparently large $n_e$ in NLRs in Type-1 AGN than 
	in Type-2 AGN.	
\item Either higher $n_e$ in NLRs in Type-1 AGN than in Type-2 AGN or expected longer time durations of AGN activities 
	triggering outflows in Type-1 AGN than in Type-2 AGN or stronger star-forming contributions in Type-1 AGN than 
	in Type-2 AGN could provide interesting challenges to the currently accepted Unified model of AGN.
\end{itemize}

\begin{table*}
\caption{Basic information of the 548 Type-1 AGN with reliable [S~{\sc ii}] emissions in the subsample}
\begin{center}
\begin{tabular}{ccccccccccc}
\hline\hline
	(1) & (2) & (3) & (4) & (5) & (6) & (7) & (8) & (9) & (10) & (11) \\
\hline
	0271-51883-0322 & 0.086 & 20.59 & 1.064 & 0.181 & 0.832 & 41.58 & 0.775 & \dots & \dots & 3.090\\
	0273-51957-0460 & 0.096 & 13.19 & 1.189 & 0.005 & 0.780 & 41.78 & 1.210 & \dots & \dots & 2.284\\
	0274-51913-0141 & 0.138 & 15.66 & 1.098 & 0.098 & 0.919 & 42.02 & 1.535 & 102.8 & 3.851 & \dots\\
	0279-51608-0392 & 0.072 & 17.36 & 0.932 & -0.175 & 0.433 & 41.22 & 1.141 & \dots & \dots & 2.435\\
	0288-52000-0088 & 0.118 & 11.83 & 1.266 & -0.051 & 0.759 & 41.91 & 0.854 & 185.1 & 3.799 & 2.844\\
	0289-51990-0056 & 0.114 & 15.73 & 0.612 & -0.178 & 0.339 & 41.38 & 1.158 & \dots & \dots & 2.400\\
	0291-51660-0451 & 0.089 & 17.19 & 0.882 & -0.234 & 0.432 & 41.41 & 1.302 & \dots & \dots & 2.024\\
	0291-51660-0470 & 0.142 & 11.35 & 1.021 & 0.104 & 1.245 & 41.65 & 1.234 & \dots & \dots & 2.226\\
	0291-51928-0474 & 0.142 & 16.93 & 0.954 & 0.093 & 0.834 & 41.92 & 1.266 & \dots & \dots & 2.139\\
	0295-51985-0255 & 0.086 & 13.41 & 0.740 & -0.220 & 0.480 & 40.97 & 1.490 & \dots & \dots & \dots\\
	0297-51959-0439 & 0.103 & 12.75 & 0.521 & -0.211 & 0.414 & 41.16 & 1.157 & \dots & \dots & 2.401\\
	0308-51662-0007 & 0.119 & 12.92 & 0.942 & -0.023 & 0.686 & 42.26 & 1.298 & \dots & \dots & 2.040\\
	0308-51662-0008 & 0.119 & 15.80 & 1.034 & -0.008 & 1.005 & 42.21 & 1.253 & 164.1 & 3.809 & 2.079\\
	0309-51666-0116 & 0.132 & 14.31 & 1.150 & -0.120 & 0.491 & 41.99 & 1.310 & 23.66 & 4.016 & 2.006\\
	0326-52375-0259 & 0.096 & 15.44 & 0.692 & -0.153 & 0.356 & 41.21 & 1.122 & \dots & \dots & 2.472\\
	0326-52375-0580 & 0.119 & 13.18 & 1.209 & 0.030 & 1.157 & 42.03 & 0.774 & \dots & \dots & 3.092\\
	0330-52370-0168 & 0.165 & 11.64 & 0.840 & 0.007 & 0.452 & 41.86 & 1.139 & \dots & \dots & 2.438\\
	0336-51999-0611 & 0.086 & 18.09 & 0.812 & -0.106 & 0.268 & 42.08 & 1.118 & \dots & \dots & 2.480\\
	0339-51692-0544 & 0.159 & 14.10 & 0.817 & -0.036 & 0.464 & 42.23 & 1.013 & \dots & \dots & 2.673\\
	0341-51690-0131 & 0.085 & 24.22 & 0.808 & -0.119 & 0.349 & 41.67 & 1.144 & \dots & \dots & 2.429\\
	0349-51699-0624 & 0.176 & 13.91 & 1.012 & -0.096 & 0.665 & 42.70 & 1.427 & \dots & \dots & 1.138\\
	0350-51691-0169 & 0.228 & 10.07 & 0.773 & -0.287 & 0.351 & 42.11 & 1.401 & \dots & \dots & 1.489\\
	0350-51691-0520 & 0.078 & 21.99 & 0.965 & -0.011 & 0.799 & 41.37 & 1.304 & 46.21 & 3.933 & 1.987\\
	0354-51792-0617 & 0.068 & 24.71 & 1.021 & 0.101 & 0.639 & 41.68 & 1.421 & 59.88 & 3.905 & 1.212\\
	0355-51788-0319 & 0.125 & 12.04 & 1.107 & 0.087 & 1.951 & 41.48 & 1.045 & \dots & \dots & 2.615\\
	0366-52017-0299 & 0.159 & 18.58 & 0.998 & -0.163 & 0.370 & 42.31 & 0.739 & \dots & \dots & 3.160\\
	0371-52078-0215 & 0.110 & 15.14 & 1.013 & -0.162 & 0.635 & 42.16 & 1.107 & 40.86 & 3.947 & 2.475\\
	0371-52078-0457 & 0.224 & 8.549 & 1.003 & -0.019 & 0.398 & 41.91 & 1.178 & \dots & \dots & 2.357\\
	0372-52173-0446 & 0.217 & 11.28 & 0.739 & -0.215 & 0.307 & 42.31 & 1.048 & \dots & \dots & 2.610\\
	0381-51811-0403 & 0.198 & 9.365 & 0.787 & -0.174 & 0.457 & 41.71 & 0.949 & \dots & \dots & 2.782\\
	0385-51783-0251 & 0.166 & 10.93 & 1.069 & -0.091 & 0.814 & 41.82 & 1.055 & 9.126 & 4.172 & 2.683\\
	0385-51783-0547 & 0.097 & 17.50 & 1.026 & -0.087 & 0.655 & 41.62 & 1.369 & 13.40 & 4.103 & 1.780\\
	0385-51877-0550 & 0.097 & 18.68 & 0.983 & -0.123 & 0.619 & 41.52 & 1.036 & \dots & \dots & 2.632\\
	0388-51793-0408 & 0.113 & 14.78 & 1.185 & -0.018 & 0.633 & 41.74 & 1.758 & 6.329 & 4.251 & \dots\\
	0390-51900-0327 & 0.063 & 24.88 & 0.749 & -0.215 & 0.409 & 41.35 & 1.039 & \dots & \dots & 2.626\\
	0400-51820-0047 & 0.098 & 21.69 & 1.058 & 0.023 & 0.599 & 42.03 & 1.047 & \dots & \dots & 2.611\\
	0404-51812-0343 & 0.078 & 18.27 & 0.735 & -0.160 & 0.708 & 41.02 & 1.078 & \dots & \dots & 2.555\\
	0404-51877-0347 & 0.078 & 16.13 & 0.716 & -0.188 & 0.657 & 41.00 & 0.928 & \dots & \dots & 2.818\\
\hline
\end{tabular}
\end{center}
\tablecomments{
The first column shows the pmf information of plate-mjd-fiberid of each Type-1 AGN, the second column shows the redshift 
information of each Type-1 AGN, the third column shows the SN information of SDSS spectrum of each Type-1 AGN, the fourth 
column to the sixth column show $\log(O3HB)$, $\log(N2HA)$ and $\log(S2HA)$ of each Type-1 AGN, the seventh column shows 
the logarithmic line luminosity (in units of erg/s) of [O~{\sc iii}]$\lambda5007$\AA~ of each Type-1 AGN, 
the eighth column shows the determined $R_{sii}$ of each Type-1 AGN, the ninth column shows the determined $O_{32}$ of each 
Type-1 AGN, the tenth column shows the determined logarithmic electron temperature $\log(T_e)$ (in units of K) of the 
Type-1 AGN with reliable $O_{32}$, the final column shows the determined logarithmic $n_e$ (in units of ${\rm cm^{-3}}$) 
of each Type-1 AGN. \\
In the last three columns, $\dots$ means no reliable value.\\
For the last column, if there is reliable $T_e$, the $n_e$ is the value after correction of effects of $T_e$, if there 
is not reliable $T_e$, the $n_e$ is the value determined through $R_{sii}$. \\
In the last column, $\dots$ means $R_{sii}$ is so large that the equation 5\ in the manuscript can not lead to a reliable value.
}
\end{table*}

\setcounter{table}{0}

\begin{table*}
\caption{--to be continued}
\begin{center}
\begin{tabular}{ccccccccccc}
\hline\hline
	(1) & (2) & (3) & (4) & (5) & (6) & (7) & (8) & (9) & (10) & (11) \\
\hline
	0404-51877-0409 & 0.180 & 10.12 & 0.872 & -0.132 & 0.651 & 42.24 & 1.080 & \dots & \dots & 2.551\\
	0406-51876-0112 & 0.213 & 8.267 & 0.908 & -0.075 & 0.370 & 42.17 & 0.839 & \dots & \dots & 2.972\\
	0412-52258-0083 & 0.281 & 10.04 & 0.754 & -0.122 & 0.466 & 42.84 & 1.263 & \dots & \dots & 2.148\\
	0412-52258-0576 & 0.208 & 9.968 & 0.677 & -0.182 & 0.315 & 41.67 & 1.538 & \dots & \dots & \dots\\
	0413-51821-0535 & 0.067 & 21.20 & 1.050 & -0.152 & 0.345 & 41.34 & 1.147 & 24.77 & 4.010 & 2.427\\
	0413-51929-0528 & 0.067 & 19.26 & 0.956 & -0.162 & 0.432 & 41.45 & 1.061 & 29.91 & 3.985 & 2.580\\
	0418-51817-0281 & 0.100 & 17.43 & 0.811 & -0.198 & 0.361 & 41.73 & 1.228 & \dots & \dots & 2.240\\
\hline
\end{tabular}
\end{center}
\end{table*}

\setcounter{table}{0}

\begin{table*}
\caption{--to be continued}
\begin{center}
\begin{tabular}{ccccccccccc}
\hline\hline
	(1) & (2) & (3) & (4) & (5) & (6) & (7) & (8) & (9) & (10) & (11) \\
\hline
	0568-52254-0184 & 0.128 & 25.12 & 1.056 & -0.079 & 0.346 & 42.46 & 0.995 & 27.09 & 3.998 & 2.702\\
	0572-52289-0521 & 0.124 & 25.28 & 0.661 & -0.178 & 0.191 & 42.00 & 1.248 & \dots & \dots & 2.187\\
	0575-52319-0202 & 0.061 & 26.13 & 0.932 & -0.076 & 0.405 & 41.61 & 1.190 & 32.82 & 3.974 & 2.317\\
	0576-52325-0413 & 0.122 & 14.03 & 0.646 & -0.226 & 0.397 & 41.19 & 1.301 & 2.968 & 4.480 & 2.269\\
	0580-52368-0532 & 0.095 & 14.44 & 1.170 & -0.122 & 0.691 & 42.08 & 1.004 & \dots & \dots & 2.688\\
	0581-52356-0418 & 0.193 & 11.54 & 0.486 & -0.187 & 0.313 & 42.12 & 0.996 & \dots & \dots & 2.702\\
	0583-52055-0341 & 0.093 & 17.74 & 0.751 & -0.145 & 0.343 & 42.00 & 1.150 & \dots & \dots & 2.417\\
\hline
\end{tabular}
\end{center}
\end{table*}

\setcounter{table}{0}

\begin{table*}
\caption{--to be continued}
\begin{center}
\begin{tabular}{ccccccccccc}
	\hline\hline
(1) & (2) & (3) & (4) & (5) & (6) & (7) & (8) & (9) & (10) & (11) \\
\hline
	0755-52235-0384 & 0.118 & 13.86 & 0.452 & -0.185 & 0.344 & 41.18 & 1.090 & \dots & \dots & 2.533\\
	0755-52235-0597 & 0.150 & 17.57 & 0.460 & -0.164 & 0.375 & 41.65 & 0.966 & \dots & \dots & 2.752\\
	0756-52577-0376 & 0.151 & 10.98 & 0.954 & -0.077 & 0.637 & 41.83 & 1.715 & \dots & \dots & \dots\\
	0756-52577-0504 & 0.112 & 17.75 & 0.571 & -0.157 & 0.303 & 41.43 & 1.365 & \dots & \dots & 1.751\\
	0759-52254-0070 & 0.079 & 17.66 & 0.730 & -0.149 & 0.352 & 41.47 & 1.117 & \dots & \dots & 2.481\\
	0762-52232-0009 & 0.126 & 15.17 & 0.480 & -0.224 & 0.326 & 41.41 & 1.133 & \dots & \dots & 2.450\\
	0765-52254-0518 & 0.141 & 14.98 & 0.523 & -0.264 & 0.305 & 41.53 & 0.944 & \dots & \dots & 2.791\\
\hline
\end{tabular}
\end{center}
\end{table*}

\setcounter{table}{0}

\begin{table*}
\caption{--to be continued}
\begin{center}
\begin{tabular}{ccccccccccc}
\hline\hline
(1) & (2) & (3) & (4) & (5) & (6) & (7) & (8) & (9) & (10) & (11) \\
\hline
	0914-52721-0410 & 0.145 & 14.30 & 0.951 & -0.010 & 0.689 & 41.76 & 1.133 & \dots & \dots & 2.451\\
	0919-52409-0173 & 0.150 & 9.202 & 0.871 & -0.137 & 0.397 & 41.57 & 1.071 & \dots & \dots & 2.569\\
	0919-52409-0487 & 0.128 & 13.54 & 0.962 & 0.052 & 0.561 & 41.82 & 1.124 & 51.22 & 3.922 & 2.429\\
	0919-52409-0619 & 0.138 & 16.22 & 0.603 & -0.052 & 0.331 & 41.94 & 1.229 & \dots & \dots & 2.237\\
	0920-52411-0329 & 0.138 & 14.38 & 0.615 & -0.043 & 0.342 & 41.90 & 1.038 & \dots & \dots & 2.627\\
	0921-52380-0244 & 0.085 & 10.71 & 0.665 & -0.108 & 0.655 & 41.16 & 1.136 & \dots & \dots & 2.444\\
	0923-52404-0586 & 0.119 & 18.08 & 0.777 & -0.144 & 0.715 & 41.95 & 1.041 & \dots & \dots & 2.623\\
\hline
\end{tabular}
\end{center}
\end{table*}

\setcounter{table}{0}

\begin{table*}
\caption{--to be continued}
\begin{center}
\begin{tabular}{ccccccccccc}
\hline\hline
(1) & (2) & (3) & (4) & (5) & (6) & (7) & (8) & (9) & (10) & (11) \\
\hline
	1058-52520-0327 & 0.150 & 12.38 & 0.960 & -0.073 & 0.571 & 41.91 & 1.163 & \dots & \dots & 2.389\\
	1061-52641-0095 & 0.097 & 10.49 & 0.879 & -0.090 & 0.461 & 41.39 & 1.016 & \dots & \dots & 2.666\\
	1078-52643-0571 & 0.124 & 13.26 & 0.793 & -0.229 & 0.557 & 41.44 & 1.238 & \dots & \dots & 2.214\\
	1158-52668-0190 & 0.094 & 16.34 & 1.222 & -0.238 & 0.358 & 41.79 & 1.004 & \dots & \dots & 2.688\\
	1162-52668-0063 & 0.104 & 14.19 & 0.921 & -0.368 & 0.392 & 41.46 & 1.306 & \dots & \dots & 2.012\\
	1163-52669-0036 & 0.080 & 13.45 & 0.584 & -0.250 & 0.315 & 41.38 & 1.290 & \dots & \dots & 2.067\\
	1170-52756-0567 & 0.138 & 13.76 & 1.154 & -0.064 & 0.606 & 42.13 & 1.247 & \dots & \dots & 2.191\\
\hline
\end{tabular}
\end{center}
\end{table*}

\setcounter{table}{0}

\begin{table*}
\caption{--to be continued}
\begin{center}
\begin{tabular}{ccccccccccc}
\hline\hline
(1) & (2) & (3) & (4) & (5) & (6) & (7) & (8) & (9) & (10) & (11) \\
\hline
	1358-52994-0218 & 0.085 & 16.90 & 1.169 & -0.096 & 0.449 & 41.78 & 0.973 & \dots & \dots & 2.741\\
	1363-53053-0274 & 0.074 & 27.55 & 0.872 & -0.143 & 0.444 & 41.48 & 0.892 & \dots & \dots & 2.879\\
	1367-53083-0379 & 0.115 & 24.99 & 0.797 & -0.166 & 0.351 & 42.11 & 1.139 & 16.91 & 4.065 & 2.472\\
	1368-53084-0071 & 0.087 & 20.48 & 1.117 & 0.105 & 0.986 & 41.89 & 1.162 & 116.9 & 3.839 & 2.311\\
	1372-53062-0412 & 0.030 & 30.17 & 1.130 & -0.100 & 1.104 & 40.97 & 1.098 & 46.24 & 3.933 & 2.485\\
	1375-53084-0330 & 0.182 & 8.870 & 0.820 & -0.189 & 0.515 & 41.81 & 1.127 & \dots & \dots & 2.463\\
	1375-53084-0424 & 0.060 & 19.36 & 1.152 & -0.296 & 0.332 & 41.63 & 1.111 & 98.39 & 3.855 & 2.421\\
\hline
\end{tabular}
\end{center}
\end{table*}

\setcounter{table}{0}

\begin{table*}
\caption{--to be continued}
\begin{center}
\begin{tabular}{ccccccccccc}
\hline\hline
(1) & (2) & (3) & (4) & (5) & (6) & (7) & (8) & (9) & (10) & (11) \\
\hline
	1645-53172-0292 & 0.175 & 9.499 & 0.875 & -0.148 & 0.459 & 41.97 & 0.983 & \dots & \dots & 2.724\\
	1649-53149-0044 & 0.112 & 13.35 & 0.709 & -0.271 & 0.387 & 41.31 & 1.190 & \dots & \dots & 2.331\\
	1656-53533-0141 & 0.138 & 12.93 & 1.054 & 0.024 & 0.676 & 42.14 & 1.022 & \dots & \dots & 2.656\\
	1658-53240-0627 & 0.183 & 8.698 & 0.837 & -0.089 & 0.342 & 41.78 & 1.643 & \dots & \dots & \dots\\
	1666-52991-0575 & 0.051 & 30.37 & 0.898 & -0.004 & 0.747 & 41.61 & 1.185 & \dots & \dots & 2.342\\
	1668-53433-0309 & 0.063 & 17.57 & 1.097 & -0.064 & 0.878 & 41.67 & 0.956 & \dots & \dots & 2.770\\
	1672-53460-0144 & 0.072 & 25.04 & 0.471 & -0.156 & 0.338 & 41.18 & 1.171 & \dots & \dots & 2.372\\
\hline
\end{tabular}
\end{center}
\end{table*}

\setcounter{table}{0}

\begin{table*}
\caption{--to be continued}
\begin{center}
\begin{tabular}{ccccccccccc}
\hline\hline
(1) & (2) & (3) & (4) & (5) & (6) & (7) & (8) & (9) & (10) & (11) \\
\hline
	1845-54144-0624 & 0.113 & 18.70 & 0.453 & -0.276 & 0.282 & 41.19 & 1.214 & \dots & \dots & 2.274\\
	1848-54180-0278 & 0.235 & 11.91 & 0.629 & -0.223 & 0.268 & 42.37 & 0.813 & \dots & \dots & 3.019\\
	1852-53534-0326 & 0.088 & 9.670 & 0.945 & -0.082 & 0.433 & 41.15 & 1.194 & \dots & \dots & 2.321\\
	1867-53317-0344 & 0.097 & 13.16 & 0.863 & -0.186 & 0.415 & 41.36 & 0.657 & \dots & \dots & 3.342\\
	1868-53318-0592 & 0.062 & 27.16 & 1.209 & 0.148 & 1.434 & 41.74 & 1.277 & 146.5 & 3.819 & 2.015\\
	1877-54464-0252 & 0.078 & 16.41 & 1.039 & -0.051 & 0.838 & 41.91 & 1.167 & \dots & \dots & 2.381\\
	1877-54464-0322 & 0.140 & 10.02 & 0.819 & -0.134 & 0.952 & 41.84 & 1.052 & \dots & \dots & 2.603\\
\hline
\end{tabular}
\end{center}
\end{table*}

\setcounter{table}{0}

\begin{table*}
\caption{--to be continued}
\begin{center}
\begin{tabular}{ccccccccccc}
\hline\hline
(1) & (2) & (3) & (4) & (5) & (6) & (7) & (8) & (9) & (10) & (11) \\
\hline
	2143-54184-0070 & 0.151 & 17.92 & 1.072 & -0.100 & 0.468 & 42.50 & 1.086 & \dots & \dots & 2.541\\
	2143-54184-0605 & 0.184 & 7.931 & 0.928 & -0.060 & 0.569 & 41.99 & 1.311 & \dots & \dots & 1.993\\
	2144-53770-0183 & 0.105 & 18.09 & 0.897 & 0.029 & 0.401 & 41.64 & 1.053 & \dots & \dots & 2.601\\
	2145-54212-0115 & 0.174 & 12.67 & 0.765 & -0.186 & 0.332 & 41.83 & 0.996 & \dots & \dots & 2.701\\
	2145-54212-0302 & 0.060 & 16.24 & 0.859 & -0.111 & 0.589 & 41.33 & 0.779 & \dots & \dots & 3.082\\
	2149-54509-0069 & 0.063 & 20.81 & 1.163 & 0.023 & 0.840 & 41.47 & 0.929 & 44.56 & 3.937 & 2.785\\
	2150-54510-0283 & 0.083 & 25.88 & 0.917 & -0.093 & 0.401 & 41.50 & 0.979 & 35.78 & 3.963 & 2.712\\
\hline
\end{tabular}
\end{center}
\end{table*}

\setcounter{table}{0}

\begin{table*}
\caption{--to be continued}
\begin{center}
\begin{tabular}{ccccccccccc}
\hline\hline
(1) & (2) & (3) & (4) & (5) & (6) & (7) & (8) & (9) & (10) & (11) \\
\hline
	2497-54154-0007 & 0.117 & 19.88 & 0.903 & -0.241 & 0.389 & 42.26 & 1.063 & \dots & \dots & 2.583\\
	2499-54176-0555 & 0.157 & 14.59 & 0.738 & -0.197 & 0.470 & 41.82 & 0.975 & \dots & \dots & 2.737\\
	2500-54178-0197 & 0.079 & 22.02 & 0.934 & -0.200 & 0.398 & 41.33 & 1.030 & 40.19 & 3.949 & 2.617\\
	2500-54178-0327 & 0.102 & 15.89 & 1.191 & 0.095 & 0.813 & 41.90 & 1.015 & \dots & \dots & 2.668\\
	2507-53876-0416 & 0.035 & 26.24 & 1.108 & -0.079 & 0.552 & 41.13 & 1.097 & 26.61 & 4.001 & 2.520\\
	2507-53876-0424 & 0.209 & 9.439 & 0.558 & -0.306 & 0.262 & 41.74 & 1.107 & \dots & \dots & 2.501\\
	2509-54180-0303 & 0.095 & 16.49 & 0.544 & -0.070 & 0.491 & 41.08 & 1.280 & \dots & \dots & 2.098\\
\hline
\end{tabular}
\end{center}
\end{table*}

\setcounter{table}{0}

\begin{table*}
\caption{--to be continued}
\begin{center}
\begin{tabular}{ccccccccccc}
\hline\hline
(1) & (2) & (3) & (4) & (5) & (6) & (7) & (8) & (9) & (10) & (11) \\
\hline
	2767-54243-0077 & 0.149 & 14.95 & 1.074 & -0.013 & 0.888 & 42.39 & 1.067 & \dots & \dots & 2.576\\
	2767-54243-0278 & 0.142 & 13.22 & 1.131 & 0.165 & 0.967 & 42.27 & 1.517 & \dots & \dots & \dots\\
	2769-54527-0624 & 0.064 & 19.01 & 0.857 & -0.254 & 0.359 & 41.15 & 1.120 & \dots & \dots & 2.476\\
	2782-54592-0070 & 0.070 & 19.98 & 0.870 & -0.144 & 0.413 & 41.47 & 1.119 & \dots & \dots & 2.478\\
	2784-54529-0142 & 0.184 & 14.28 & 0.842 & -0.072 & 0.454 & 41.90 & 1.311 & \dots & \dots & 1.995\\
	2792-54556-0606 & 0.073 & 20.24 & 1.125 & 0.218 & 1.749 & 41.55 & 0.904 & \dots & \dots & 2.858\\
	2881-54502-0432 & 0.111 & 15.69 & 1.207 & 0.090 & 0.530 & 42.18 & 1.022 & \dots & \dots & 2.656\\
	2886-54498-0353 & 0.117 & 20.90 & 1.063 & 0.060 & 0.615 & 41.86 & 0.976 & \dots & \dots & 2.735\\
	2947-54533-0076 & 0.134 & 15.20 & 0.681 & -0.076 & 0.419 & 41.38 & 1.204 & \dots & \dots & 2.298\\
	2952-54559-0406 & 0.124 & 21.73 & 0.963 & -0.117 & 0.285 & 42.07 & 1.193 & \dots & \dots & 2.323\\
	2973-54591-0085 & 0.040 & 28.63 & 0.916 & -0.219 & 0.290 & 41.32 & 1.321 & \dots & \dots & 1.954\\
	2973-54591-0101 & 0.089 & 15.24 & 1.095 & -0.040 & 0.708 & 41.58 & 1.107 & \dots & \dots & 2.501\\
\hline
\end{tabular}
\end{center}
\end{table*}

\begin{table*}
\caption{Basic information of the 548 Type-2 AGN with reliable [S~{\sc ii}] emissions in the subsample}
\begin{center}
\begin{tabular}{ccccccccccc}
\hline\hline
(1) & (2) & (3) & (4) & (5) & (6) & (7) & (8) & (9) & (10) & (11) \\
\hline
	0271-51883-0520 & 0.092 & 16.45 & 1.123 & 0.151 & 0.569 & 41.88 & 1.271 & \dots & \dots & 2.124\\
	0272-51941-0429 & 0.167 & 8.227 & 0.804 & -0.212 & 0.395 & 41.81 & 1.169 & \dots & \dots & 2.376\\
	0272-51941-0529 & 0.177 & 12.20 & 1.107 & -0.004 & 0.645 & 42.34 & 1.087 & \dots & \dots & 2.539\\
	0274-51913-0115 & 0.095 & 17.53 & 0.688 & -0.213 & 0.550 & 40.91 & 1.175 & \dots & \dots & 2.364\\
	0275-51910-0115 & 0.127 & 14.54 & 0.459 & -0.250 & 0.515 & 40.98 & 0.693 & \dots & \dots & 3.256\\
	0276-51909-0147 & 0.108 & 17.80 & 0.452 & -0.261 & 0.542 & 41.04 & 1.130 & \dots & \dots & 2.456\\
	0277-51908-0189 & 0.068 & 16.78 & 0.950 & -0.012 & 0.704 & 41.01 & 1.217 & \dots & \dots & 2.268\\
\hline
\end{tabular}
\end{center}
	\tablecomments{Column information are similar as those in Table~1, but for Type-2 AGN in the subsample.}
\end{table*}

\setcounter{table}{1}

\begin{table*}
\caption{--to be continued}
\begin{center}\begin{tabular}{ccccccccccc}
\hline\hline
(1) & (2) & (3) & (4) & (5) & (6) & (7) & (8) & (9) & (10) & (11) \\
\hline
	0412-52258-0104 & 0.074 & 21.29 & 0.952 & -0.176 & 0.602 & 41.56 & 1.265 & 64.08 & 3.898 & 2.089\\
	0414-51869-0027 & 0.149 & 12.06 & 1.056 & 0.130 & 0.826 & 41.66 & 1.438 & \dots & \dots & \dots\\
	0414-51869-0257 & 0.117 & 18.73 & 0.476 & -0.250 & 0.418 & 41.22 & 1.533 & \dots & \dots & \dots\\
	0414-51901-0188 & 0.066 & 25.17 & 1.051 & -0.007 & 0.916 & 41.54 & 0.970 & \dots & \dots & 2.745\\
	0416-51885-0279 & 0.253 & 6.730 & 0.651 & -0.130 & 0.259 & 41.69 & 0.829 & \dots & \dots & 2.990\\
	0417-51821-0264 & 0.094 & 16.06 & 0.863 & -0.227 & 0.440 & 41.49 & 1.271 & \dots & \dots & 2.124\\
	0419-51868-0383 & 0.102 & 12.67 & 0.521 & -0.040 & 0.471 & 41.35 & 1.281 & 49.38 & 3.926 & 2.058\\
\hline\end{tabular}\end{center}
\end{table*}

\setcounter{table}{1}

\begin{table*}
\caption{--to be continued}
\begin{center}\begin{tabular}{ccccccccccc}
\hline\hline
(1) & (2) & (3) & (4) & (5) & (6) & (7) & (8) & (9) & (10) & (11) \\
\hline
	0601-52316-0027 & 0.201 & 9.931 & 0.716 & -0.174 & 0.403 & 41.59 & 0.975 & \dots & \dots & 2.737\\
	0604-52079-0490 & 0.123 & 10.21 & 1.103 & -0.059 & 0.529 & 41.95 & 1.172 & 93.84 & 3.859 & 2.299\\
	0604-52079-0640 & 0.200 & 9.269 & 0.791 & -0.166 & 0.594 & 41.57 & 1.132 & \dots & \dots & 2.453\\
	0606-52365-0535 & 0.174 & 16.31 & 0.752 & -0.109 & 0.435 & 41.83 & 1.166 & \dots & \dots & 2.383\\
	0607-52368-0572 & 0.187 & 14.80 & 0.828 & -0.007 & 0.579 & 42.08 & 1.128 & \dots & \dots & 2.460\\
	0607-52368-0637 & 0.186 & 11.57 & 0.656 & -0.191 & 0.291 & 41.78 & 1.615 & 31.83 & 3.978 & \dots\\
	0610-52056-0259 & 0.158 & 7.670 & 0.802 & -0.174 & 0.531 & 41.76 & 0.866 & \dots & \dots & 2.924\\
\hline\end{tabular}\end{center}
\end{table*}

\setcounter{table}{1}

\begin{table*}
\caption{--to be continued}
\begin{center}\begin{tabular}{ccccccccccc}
\hline\hline
(1) & (2) & (3) & (4) & (5) & (6) & (7) & (8) & (9) & (10) & (11) \\
\hline
	0769-54530-0225 & 0.038 & 24.33 & 0.815 & -0.295 & 0.556 & 40.66 & 1.146 & 97.36 & 3.856 & 2.353\\
	0781-52373-0136 & 0.102 & 12.68 & 0.488 & -0.190 & 0.307 & 41.13 & 1.079 & \dots & \dots & 2.553\\
	0784-52327-0314 & 0.115 & 18.04 & 0.870 & -0.057 & 0.440 & 41.58 & 1.381 & \dots & \dots & 1.656\\
	0785-52339-0368 & 0.090 & 14.68 & 0.837 & -0.134 & 0.527 & 41.66 & 1.268 & 53.17 & 3.918 & 2.092\\
	0786-52319-0568 & 0.073 & 15.88 & 1.078 & -0.187 & 0.428 & 41.32 & 1.112 & 62.11 & 3.901 & 2.442\\
	0788-52338-0398 & 0.097 & 18.22 & 0.755 & -0.194 & 0.589 & 41.45 & 1.163 & \dots & \dots & 2.388\\
	0790-52433-0346 & 0.154 & 8.299 & 0.599 & -0.203 & 0.413 & 42.02 & 1.176 & \dots & \dots & 2.361\\
\hline\end{tabular}\end{center}
\end{table*}

\setcounter{table}{1}

\begin{table*}
\caption{--to be continued}
\begin{center}\begin{tabular}{ccccccccccc}
\hline\hline
(1) & (2) & (3) & (4) & (5) & (6) & (7) & (8) & (9) & (10) & (11) \\
\hline
	0931-52619-0041 & 0.170 & 10.13 & 1.070 & -0.076 & 0.445 & 41.81 & 1.320 & \dots & \dots & 1.959\\
	0931-52619-0094 & 0.097 & 23.72 & 1.046 & 0.005 & 0.851 & 41.87 & 1.248 & \dots & \dots & 2.189\\
	0931-52619-0613 & 0.109 & 15.08 & 0.952 & -0.149 & 0.355 & 41.87 & 1.269 & \dots & \dots & 2.131\\
	0933-52642-0116 & 0.084 & 17.71 & 0.701 & -0.139 & 0.651 & 41.21 & 1.311 & \dots & \dots & 1.994\\
	0936-52705-0183 & 0.104 & 13.96 & 1.056 & -0.138 & 0.742 & 41.60 & 1.278 & \dots & \dots & 2.103\\
	0940-52670-0043 & 0.103 & 16.22 & 1.051 & 0.080 & 0.646 & 41.92 & 1.206 & \dots & \dots & 2.293\\
	0940-52670-0101 & 0.071 & 28.17 & 0.740 & -0.125 & 0.367 & 42.19 & 1.315 & \dots & \dots & 1.978\\
\hline\end{tabular}\end{center}
\end{table*}

\setcounter{table}{1}

\begin{table*}
\caption{--to be continued}
\begin{center}\begin{tabular}{ccccccccccc}
\hline\hline
(1) & (2) & (3) & (4) & (5) & (6) & (7) & (8) & (9) & (10) & (11) \\
\hline
	1197-52668-0459 & 0.136 & 16.23 & 0.795 & -0.066 & 0.461 & 41.56 & 1.293 & \dots & \dots & 2.057\\
	1202-52672-0120 & 0.128 & 13.52 & 1.133 & -0.084 & 0.328 & 42.00 & 1.601 & 46.19 & 3.933 & \dots\\
	1211-52964-0023 & 0.183 & 6.492 & 0.715 & -0.128 & 0.304 & 41.64 & 1.142 & \dots & \dots & 2.432\\
	1226-52734-0132 & 0.068 & 20.16 & 0.905 & -0.074 & 0.498 & 41.69 & 1.192 & \dots & \dots & 2.327\\
	1227-52733-0481 & 0.175 & 9.231 & 0.398 & -0.268 & 0.363 & 41.51 & 1.087 & \dots & \dots & 2.538\\
	1235-52734-0296 & 0.068 & 22.74 & 1.133 & 0.125 & 0.955 & 41.43 & 1.261 & \dots & \dots & 2.153\\
	1240-52734-0597 & 0.077 & 23.04 & 0.885 & -0.188 & 0.491 & 41.24 & 1.182 & \dots & \dots & 2.348\\
\hline\end{tabular}\end{center}
\end{table*}

\setcounter{table}{1}

\begin{table*}
\caption{--to be continued}
\begin{center}\begin{tabular}{ccccccccccc}
\hline\hline
(1) & (2) & (3) & (4) & (5) & (6) & (7) & (8) & (9) & (10) & (11) \\
\hline
	1459-53117-0040 & 0.193 & 11.89 & 0.437 & -0.151 & 0.361 & 42.14 & 1.708 & \dots & \dots & \dots\\
	1462-53112-0395 & 0.199 & 7.570 & 0.823 & -0.108 & 0.586 & 41.93 & 1.390 & \dots & \dots & 1.586\\
	1465-53082-0345 & 0.123 & 16.39 & 0.461 & -0.256 & 0.290 & 41.64 & 1.163 & \dots & \dots & 2.390\\
	1495-52944-0248 & 0.130 & 14.42 & 0.481 & -0.198 & 0.421 & 41.32 & 1.036 & \dots & \dots & 2.631\\
	1501-53740-0460 & 0.173 & 9.414 & 0.700 & -0.278 & 0.322 & 41.54 & 1.156 & \dots & \dots & 2.403\\
	1541-53732-0123 & 0.154 & 10.25 & 1.030 & -0.024 & 0.549 & 41.66 & 1.496 & \dots & \dots & \dots\\
	1542-53734-0101 & 0.166 & 12.00 & 0.441 & -0.124 & 0.661 & 41.33 & 1.489 & \dots & \dots & \dots\\
\hline\end{tabular}\end{center}
\end{table*}

\setcounter{table}{1}

\begin{table*}
\caption{--to be continued}
\begin{center}\begin{tabular}{ccccccccccc}
\hline\hline
(1) & (2) & (3) & (4) & (5) & (6) & (7) & (8) & (9) & (10) & (11) \\
\hline
	1707-53885-0227 & 0.221 & 8.461 & 0.961 & -0.028 & 0.639 & 42.01 & 0.932 & \dots & \dots & 2.811\\
	1709-53533-0241 & 0.132 & 16.33 & 0.884 & -0.070 & 0.548 & 41.85 & 1.344 & \dots & \dots & 1.859\\
	1712-53531-0131 & 0.128 & 12.72 & 1.104 & -0.018 & 0.448 & 41.95 & 1.170 & \dots & \dots & 2.374\\
	1712-53531-0482 & 0.119 & 13.00 & 0.618 & -0.114 & 0.471 & 41.33 & 1.380 & \dots & \dots & 1.662\\
	1715-54212-0530 & 0.168 & 10.35 & 1.169 & -0.126 & 0.562 & 42.21 & 1.296 & \dots & \dots & 2.045\\
	1719-53876-0580 & 0.123 & 12.65 & 0.891 & -0.092 & 0.395 & 41.72 & 1.629 & \dots & \dots & \dots\\
	1722-53852-0071 & 0.131 & 20.64 & 0.717 & -0.071 & 0.616 & 41.36 & 1.361 & \dots & \dots & 1.775\\
\hline\end{tabular}\end{center}
\end{table*}

\setcounter{table}{1}

\begin{table*}
\caption{--to be continued}
\begin{center}\begin{tabular}{ccccccccccc}
\hline\hline
(1) & (2) & (3) & (4) & (5) & (6) & (7) & (8) & (9) & (10) & (11) \\
\hline
	1945-53387-0487 & 0.133 & 12.92 & 0.797 & -0.291 & 0.486 & 41.70 & 1.009 & \dots & \dots & 2.678\\
	1945-53387-0506 & 0.079 & 25.07 & 0.682 & -0.201 & 0.350 & 41.29 & 1.248 & \dots & \dots & 2.189\\
	1946-53432-0368 & 0.137 & 11.87 & 0.835 & -0.125 & 0.409 & 41.65 & 1.242 & \dots & \dots & 2.206\\
	1948-53388-0376 & 0.144 & 10.01 & 0.870 & -0.097 & 0.553 & 41.59 & 1.317 & \dots & \dots & 1.970\\
	1959-53440-0152 & 0.141 & 10.19 & 0.750 & -0.183 & 0.375 & 41.66 & 2.035 & \dots & \dots & \dots\\
	1959-53440-0206 & 0.137 & 14.88 & 0.607 & -0.083 & 0.428 & 41.87 & 1.097 & \dots & \dots & 2.519\\
	1972-53466-0525 & 0.179 & 12.78 & 0.931 & -0.006 & 0.333 & 42.27 & 1.349 & \dots & \dots & 1.835\\
\hline\end{tabular}\end{center}
\end{table*}

\setcounter{table}{1}

\begin{table*}
\caption{--to be continued}
\begin{center}\begin{tabular}{ccccccccccc}
\hline\hline
(1) & (2) & (3) & (4) & (5) & (6) & (7) & (8) & (9) & (10) & (11) \\
\hline
	2216-53795-0209 & 0.050 & 19.70 & 0.805 & -0.060 & 0.600 & 41.23 & 1.148 & \dots & \dots & 2.421\\
	2217-53794-0566 & 0.146 & 13.73 & 0.850 & 0.124 & 0.774 & 41.70 & 1.366 & \dots & \dots & 1.747\\
	2219-53816-0090 & 0.132 & 8.967 & 0.660 & -0.247 & 0.322 & 41.48 & 1.528 & \dots & \dots & \dots\\
	2219-53816-0496 & 0.033 & 31.45 & 0.917 & -0.203 & 0.474 & 41.31 & 1.170 & 110.4 & 3.844 & 2.296\\
	2220-53795-0124 & 0.172 & 13.83 & 1.090 & 0.070 & 0.557 & 42.40 & 1.104 & \dots & \dots & 2.507\\
	2226-53819-0247 & 0.156 & 15.53 & 0.853 & -0.065 & 0.544 & 42.28 & 1.342 & \dots & \dots & 1.867\\
	2230-53799-0498 & 0.106 & 18.17 & 1.196 & 0.128 & 1.277 & 41.61 & 1.342 & \dots & \dots & 1.869\\
\hline\end{tabular}\end{center}
\end{table*}

\setcounter{table}{1}

\begin{table*}
\caption{--to be continued}
\begin{center}\begin{tabular}{ccccccccccc}
\hline\hline
(1) & (2) & (3) & (4) & (5) & (6) & (7) & (8) & (9) & (10) & (11) \\
\hline
	2529-54585-0125 & 0.216 & 8.508 & 0.786 & -0.102 & 0.350 & 42.35 & 1.182 & \dots & \dots & 2.348\\
	2529-54585-0479 & 0.152 & 12.21 & 1.085 & -0.027 & 0.418 & 42.01 & 1.311 & \dots & \dots & 1.994\\
	2530-53881-0463 & 0.204 & 11.56 & 0.660 & -0.154 & 0.398 & 41.57 & 1.339 & \dots & \dots & 1.880\\
	2578-54093-0047 & 0.157 & 10.60 & 0.567 & -0.119 & 0.396 & 41.48 & 1.264 & \dots & \dots & 2.144\\
	2579-54068-0015 & 0.174 & 11.86 & 0.901 & 0.074 & 0.791 & 42.10 & 1.171 & \dots & \dots & 2.373\\
	2579-54068-0461 & 0.208 & 7.977 & 0.680 & -0.273 & 0.294 & 41.86 & 1.221 & \dots & \dots & 2.257\\
	2580-54092-0315 & 0.165 & 13.21 & 0.857 & -0.072 & 0.836 & 41.15 & 1.373 & \dots & \dots & 1.704\\
\hline\end{tabular}\end{center}
\end{table*}

\section*{Acknowledgements}

Zhang gratefully acknowledge the anonymous referee for giving us constructive and valuable comments and 
suggestions to greatly improve the paper. Zhang gratefully thanks the kind financial support from NSFC-12173020. This 
manuscript has made use of the data from the SDSS projects, \url{http://www.sdss3.org/}, managed by the Astrophysical 
Research Consortium for the Participating Institutions of the SDSS-III Collaborations. This manuscript has made use of 
the data from FIRST database \url{http://sundog.stsci.edu/}. This paper has made use of the MPFIT package 
\url{https://pages.physics.wisc.edu/~craigm/idl/cmpfit.html} and the FITEXY procedure 
\url{https://idlastro.gsfc.nasa.gov/ftp/pro/math/fitexy.pro}

\end{document}